\documentclass[preprint,12pt]{elsarticle}

\usepackage{graphicx}
\usepackage{comment}
\usepackage{amssymb}

\usepackage{lineno,hyperref}
\usepackage{amsmath,amsfonts,amsthm,amssymb,graphicx,subfigure,mathtools,tikz,hyperref,bm,blindtext,float}
\usepackage{diagbox}
\usepackage{geometry}

\journal{Journal Name}

\begin{document}

\begin{frontmatter}


\title{A composite neural network that learns from multi-fidelity data:  Application to function approximation and inverse PDE problems}



\author[brown]{Xuhui Meng}
\author[brown,PNNL]{George Em Karniadakis\corref{cor}}
\cortext[cor]{Corresponding author}
\ead[cor]{george\_karniadakis@brown.edu}

\address[brown]{Division of Applied Mathematics, Brown University, Providence, RI, USA, 02912}
\address[PNNL]{Pacific Northwest National Laboratory, Richland, WA, USA, 99354}

\begin{abstract}
Currently the training of neural networks relies on data of comparable accuracy
but in real applications only a very small set of high-fidelity data is available while inexpensive lower fidelity data may be plentiful. 
We propose a new composite neural network (NN) that can be trained based on multi-fidelity data. It is comprised of three NNs,
with the first NN trained using the low-fidelity data and coupled to two high-fidelity NNs, one with activation functions and another one without, in order to discover and exploit nonlinear and linear correlations, respectively, between the low-fidelity and the high-fidelity data. 
We first demonstrate the accuracy of the new multi-fidelity NN for approximating some standard benchmark functions but also a 20-dimensional function that is not easy to approximate with other methods, e.g. Gaussian process regression. Subsequently, we extend the  recently developed physics-informed neural networks (PINNs) to be trained with multi-fidelity data sets (MPINNs). MPINNs contain four  fully-connected neural networks, where the first one approximates the low-fidelity data, while the second and third construct the correlation between the low- and high-fidelity data and produce the multi-fidelity approximation, which is then used in the last NN that encodes the partial differential equations (PDEs). Specifically, in the two high-fidelity NNs a relaxation parameter is introduced, which can be optimized to combine the linear and nonlinear sub-networks. By optimizing this parameter, the present model is capable of learning both the linear and complex nonlinear correlations between the low- and high-fidelity data adaptively. By training the MPINNs, we can:  (1) obtain the correlation between the low- and high-fidelity data, (2) infer the quantities of interest based on a few scattered data, and (3) identify the unknown parameters in the PDEs. In particular, we employ the MPINNs to learn the hydraulic conductivity field for  unsaturated flows as well as the reactive models for reactive transport. The results demonstrate that MPINNs can achieve relatively high accuracy based on a very small set of high-fidelity data. Despite the relatively low dimension and limited number of fidelities (two-fidelity levels)  for the benchmark problems in the present study, the proposed model can be readily extended to very high-dimensional regression and classification problems involving multi-fidelity data.
\end{abstract}


\end{frontmatter}


\section{Introduction}
\label{intro}

The  recent rapid developments in deep learning have also influenced the computational modeling of physical systems, e.g. in geosciences and  engineering  \cite{alex01,bohnke2011approach,Zheng13,nguyen2013multidisciplinary}. Generally, large numbers of high-fidelity data sets are required for optimization of complex physical systems, which may lead to computationally prohibitive costs. On the other hand, inadequate high-fidelity data result in inaccurate approximations and possibly erroneous designs. Multi-fidelity modeling has been shown to be both  efficient and effective in achieving high accuracy in diverse applications by leveraging both the low- and  high-fidelity data \cite{kennedy2000predicting,forrester2007multi,raissi2016deep,raissi2017inferring}. In the framework of multi-fidelity modeling, we assume that accurate but expensive high-fidelity data are scarce, while the cheaper and less accurate low-fidelity data are abundant. An example is the use of a few experimental measurements, which are hard to obtain, combined with synthetic data obtained from running a computational model. In many cases, the low-fidelity data can supply useful information on the trends for high-fidelity data, hence multi-fidelity modeling can greatly enhance prediction accuracy based on a small set of high-fidelity data in comparison to the single-fidelity modeling \cite{kennedy2000predicting,perdikaris2017nonlinear,bonfiglio2018improving}.

The construction of  cross-correlation between the low- and high-fidelity data is crucial in multi-fidelity methods. Several methods have been developed to estimate such correlations, such as the response surface models \cite{chang1993sensitivity,vitali2002multi}, polynomial chaos expansion \cite{eldred2009recent,padron2016multi},  Gaussian process regression (GPR) \cite{forrester2007multi,raissi2017inferring,perdikaris2017nonlinear,laurenceau2008building}, artificial neural networks \cite{minisci2013robust},  and moving least squares \cite{lancaster1981surfaces,levin1998approximation}. Interested readers can refer to \cite{fernandez2016review} for a comprehensive review of these methods. Among all the existing methods, the Gaussian process regression  in combination with the linear autoregressive scheme has drawn much attention in a wide range of applications \cite{raissi2017inferring,babaee2016multi}.   For instance, Babaee {\sl et al.} applied this approach for the mixed convection to propose an improved correlation for heat transfer, which outperforms existing empirical correlation \cite{babaee2016multi}.   We note that GPR with a linear autoregressive scheme can only capture the linear correlation between the low- and high-fidelity data. Perdikaris {\sl et al.} then extended the method in \cite{kennedy2000predicting} to enable it of learning complex nonlinear correlations \cite{perdikaris2017nonlinear}; this has been successfully employed to estimate the hydraulic conductivity based on the multi-fidelity data for pressure head in subsurface flows  \cite{zheng2018adaptive}. Although great progress has already been made, the multi-fidelity approaches based on GPR still have some limitations, e.g., approximations of discontinuous functions \cite{raissi2016deep},   high-dimensional problems \cite{perdikaris2017nonlinear}, and  inverse problems with strong nonlinearities (i.e., nonlinear partial differential equations) \cite{raissi2017inferring}. In addition, optimization for GPR is  quite difficult to implement. Therefore, multi-fidelity approaches which can overcome these drawbacks are urgently needed.

Deep neural networks can easily handle  problems with almost any nonlinearities at both low- and high-dimensions. In addition, the recently proposed physics-informed neural networks (PINNs) have shown {\it expressive power} for learning the unknown parameters or functions in inverse PDE problems with  nonlinearities \cite{raissi2019physics}. Examples of successful applications of PINNs include (1) learning the velocity and pressure fields based on partial observations of spatial-temporal visualizations of a passive scalar, i.e., solute concentration \cite{raissi2018hidden},  and (2) estimation of the unknown constitutive relationship in the nonlinear diffusion equation for unsaturated flows \cite{tartakovsky2018learning}. Despite the expressive power of PINNs, it has been documented that a large set of high-fidelity data is required for identifying the unknown parameters in nonlinear PDEs. To leverage the merits of deep neural networks (DNNs) and the concept of multi-fidelity modeling, we propose to develop multi-fidelity DNNs and multi-fidelity PINNs (MPINNs), which are expected to have the following attractive features:
 (1) they can learn both the linear and nonlinear correlations adaptively; (2) they are suitable for high-dimensional problems; (3) they can handle inverse problems with strong nonlinearities; and (4) they are easy to implement, as we demonstrate in the present work.

The rest of the paper is organized as follows: the key concepts of multi-fidelity DNNs and MPINNs are presented in Sec. \ref{mpinns}, while results for function approximation and inverse PDE problems are shown in Sec. \ref{results}. Finally, a summary for this work is given in Sec. \ref{summary}. In the Appendix we include a basic review of the embedding theory.

\section{Multi-fidelity Deep Neural Networks and MPINNs}
\label{mpinns}
The key starting point in multi-fidelity modeling is to discover and exploit the relation between low- and high-fidelity data \cite{fernandez2016review}. A widely used comprehensive correlation  \cite{fernandez2016review} is expressed as 
\begin{align}\label{linear}
y_H = \rho(x) y_L + \delta(x),
\end{align}
where $y_L$ and $y_H$ are, respectively, the low- and high-fidelity data, $\rho(x)$ is the multiplicative correlation surrogate, and $\delta(x)$ is the additive correlation surrogate. It is clear that multi-fidelity models based on this relation are only capable of handling linear correlations between the two-fidelity data. However, there exist many interesting cases that go beyond the linear correlation in Eq. \eqref{linear} \cite{perdikaris2017nonlinear}. For instance, the correlation for the low-fidelity experimental data and the high-fidelity direct numerical simulations in the mixed convection flows past a cylinder is nonlinear \cite{babaee2016multi,perdikaris2017nonlinear}.   In order to capture the nonlinear correlation, we put forth a generalized autoregressive scheme, which is expressed as
\begin{align}\label{nlinear}
y_H = F(y_L) + \delta(x),
\end{align}
where $F(.)$ is an unknown (linear/nonlinear) function that maps the low-fidelity data to the high-fidelity level.  We can further write Eq. \eqref{nlinear}  as
\begin{align}\label{correlation}
y_H = \mathcal{F}(x, y_L).
\end{align}
To explore the linear/nonlinear correlation adaptively, we then decompose $\mathcal{F}(.)$  into two parts, i.e., the linear and nonlinear parts, which are expressed as 
\begin{align}
\mathcal{F} = \mathcal{F}_l + \mathcal{F}_{nl},
\end{align} 
where $\mathcal{F}_l$ and $\mathcal{F}_{nl}$  denote the linear and nonlinear terms in $\mathcal{F}$, respectively.  Now, we construct the correlation as
\begin{align}
y_H = \alpha \mathcal{F}_l(x, y_L) + (1 - \alpha) \mathcal{F}_{nl}(x, y_L) , ~ \alpha \in [0, 1],
\end{align}
where $\alpha$ is hyper-parameter to be determined by the data; its value determines the degree of
nonlinearity of the correlation between the high- and low-fidelity data.

\begin{figure}
\centering
\includegraphics[width = 0.9\textwidth]{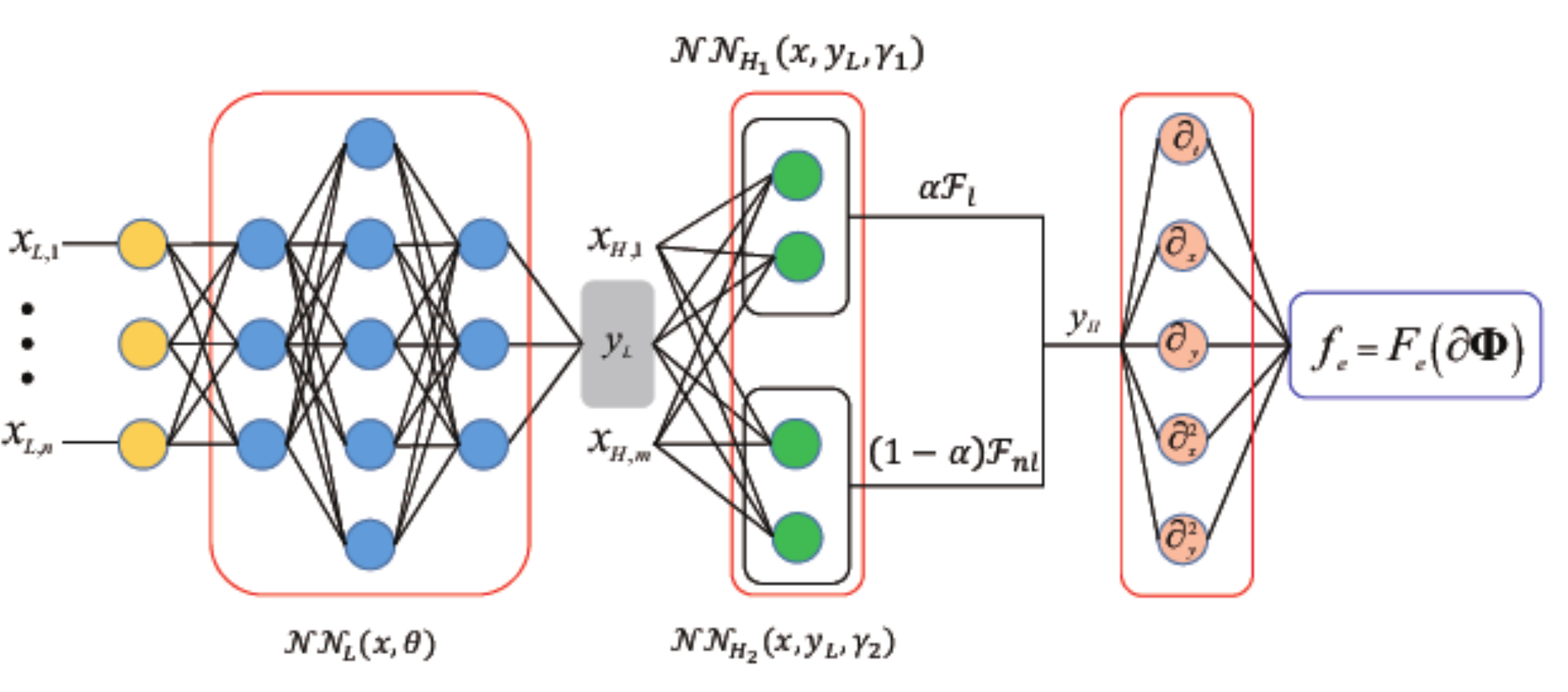}
\caption{\label{MPINNs} 
Schematic of the  multi-fidelity DNN and MPINN. The left box (blue nodes) represents the low-fidelity DNN $\mathcal{NN}_L(x,\theta)$ connected to the box with green dots representing two high fidelity DNNs, $\mathcal{NN}_{H_i}(x,y_L,\gamma_i)$ ($i=1,2$). In the case of MPINN, the combined output of the two high-fidelity DNNs is input to an additional PDE-induced DNN. Here 
$\partial \bm{\Phi} = \left[\partial_t, \partial_x, \partial_y, \partial^2_x, \partial^2_y, ... \right] y_H$ denotes symbolically the last DNN that has a very complicated graph and its structure is determined
by the specific PDE considered.
}
\end{figure}

The architecture of the proposed multi-fidelity DNN and MPINN is illustrated in Fig. \ref{MPINNs}, which is composed of four fully-connected neural networks. The first one 
$\mathcal{NN}_L(x_L, \theta)$ is employed to approximate the low-fidelity data, while the second and third NNs ($\mathcal{NN}_{H_i}(x, y_L, \beta, \gamma_i), i = 1, 2$) are for approximating the correlation for the low- and high-fidelity data; the last NN ($\mathcal{NN}_{f_e}$) is induced by encoding the governing equations, e.g. the partial differential equations (PDEs). In addition, $\mathcal{F}_l = \mathcal{NN}_{H_1}$, and $\mathcal{F}_{nl} = \mathcal{NN}_{H_2}$; $\theta$, $\alpha$, $\beta$, and $\gamma_i, i = 1, 2$ are unknown parameters of the NNs, which can be learned by minimizing the following loss function:
\begin{align}
MSE = MSE_{y_L} + MSE_{y_H} + MSE_{f_e} + \lambda \sum \beta_i^2,
\end{align}
where
\begin{align}
MSE_{y_L} &= \frac{1} {N_{y_L}} \sum^{N_{y_L}}_{i = 1}\left( |y^*_L - y_L|^2  + |\nabla y^*_L - \nabla y_L|^2 \right), \\
MSE_{y_H} &= \frac{1} {N_{y_H}} \sum^{N_{y_H}}_{i = 1}\left( |y^*_H - y_H|^2 \right),\\
MSE_{f_e} &= \frac{1} {N_{f}} \sum^{N_{f}}_{i = 1}\left( |f^*_e - f_e|^2 \right).
\end{align}
Here, $\psi$ ($\psi = y^*_L,~ y^*_H$, and $f^*_e$) denote the outputs of the $\mathcal{NN}_L$, $\mathcal{NN}_H$, and $\mathcal{NN}_{f_e}$, $\beta$ is any weight in $\mathcal{NN}_L$ and $\mathcal{NN}_{H_2}$ as well as $\alpha$, and $\lambda$ is the $L_2$ regularization rate. It is worth mentioning that the boundary/initial conditions  for $f_e$ can also be added into the loss function, in a similar fashion as in the standard PINNs introduced in detail in \cite{raissi2019physics} so we do not elaborate on this issue here. In the present study, the loss function is optimized using the L-BFGS method together with Xavier's initialization method, while the  hyperbolic tangent function is employed as the activation function in $\mathcal{NN}_L$ and $\mathcal{NN}_{H_2}$.  We note that no activation function is included in $\mathcal{NN}_{H_1}$ due to the fact that it is used to approximate the linear part of $\mathcal{F}$. Hence, the present DNNs can dynamically learn the linear as well as the nonlinear correlations without any prior knowledge on the correlation for the low- and high-fidelity data. 

\section{Results and Discussion}
\label{results}
Next we present several tests of the multi-fidelity DNN as well as the MPINN, the latter in the context of two inverse PDE problems related to geophysical applications. 

\subsection{Function approximation}
We first demonstrate the effectiveness of this multi-fidelity modeling  in approximating both continuous and discontinuous functions based on both linear and complicated nonlinear correlations between the low- and high-fidelity data.

\subsubsection{Continuous function with linear correlation}
\begin{figure}
\centering
\subfigure[]{\label{caseIa}
\includegraphics[width = 0.45\textwidth]{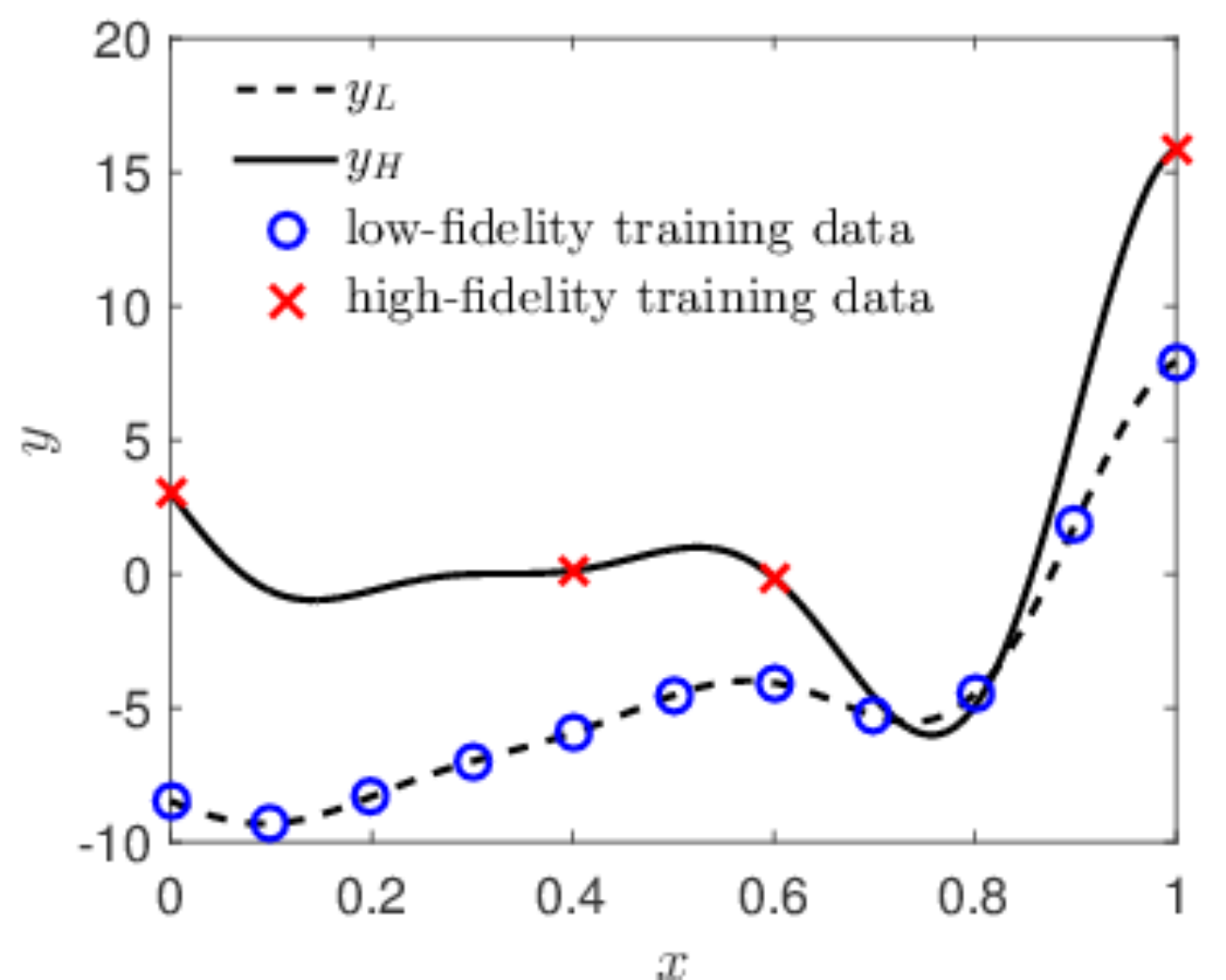}}
\subfigure[]{\label{caseIb}
\includegraphics[width = 0.45\textwidth]{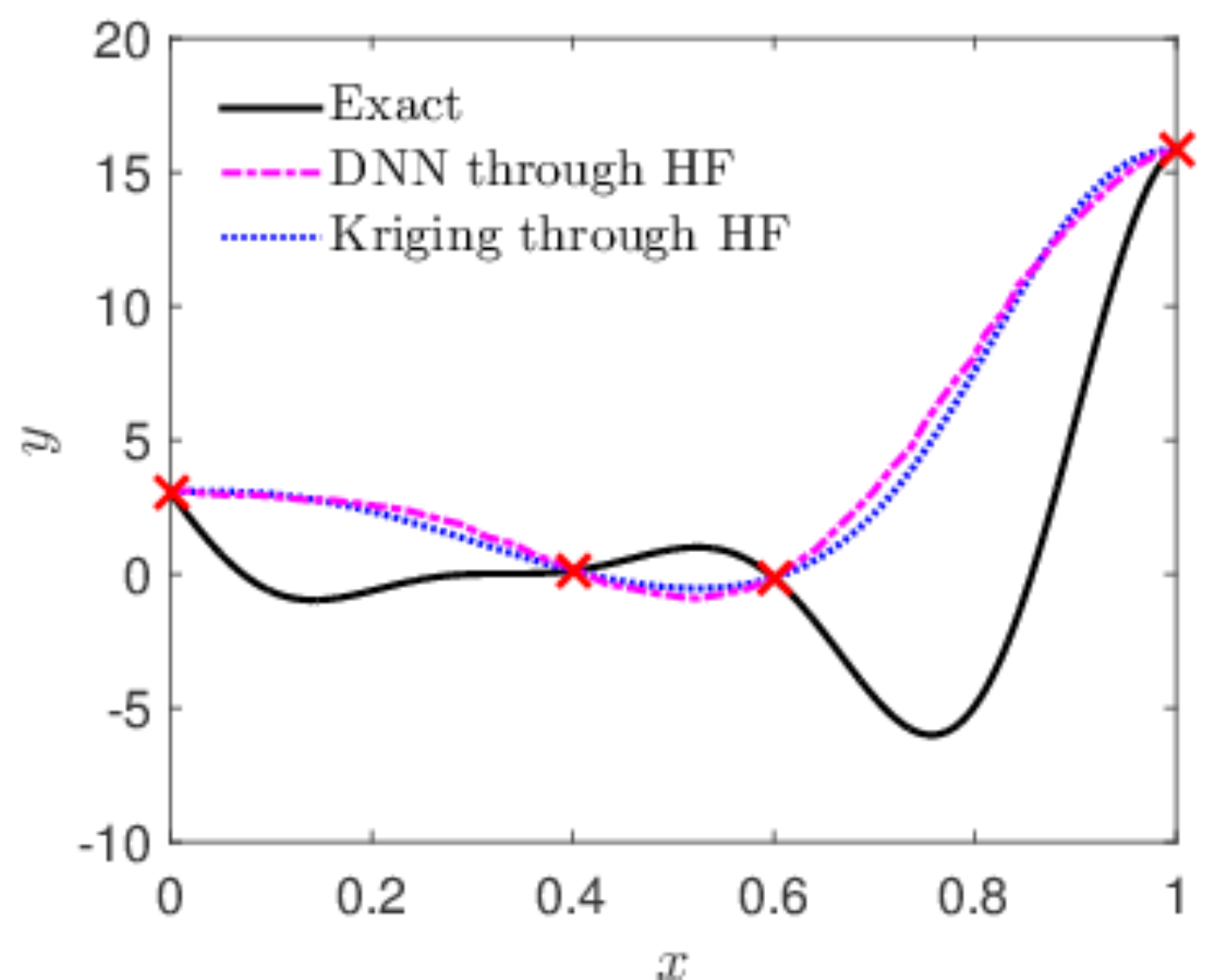}}
\subfigure[]{\label{caseIc}
\includegraphics[width = 0.45\textwidth]{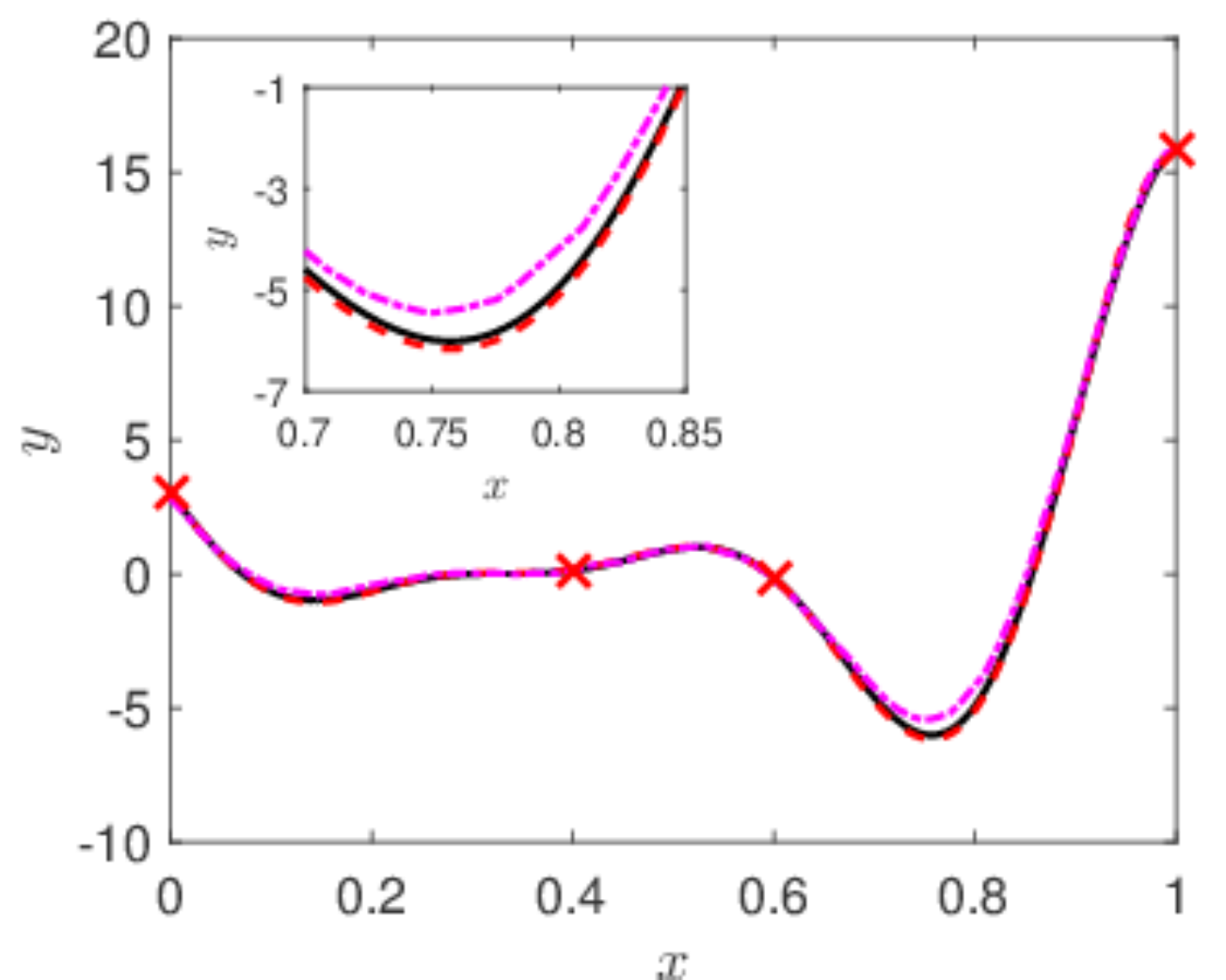}}
\subfigure[]{\label{caseId}
\includegraphics[width = 0.45\textwidth]{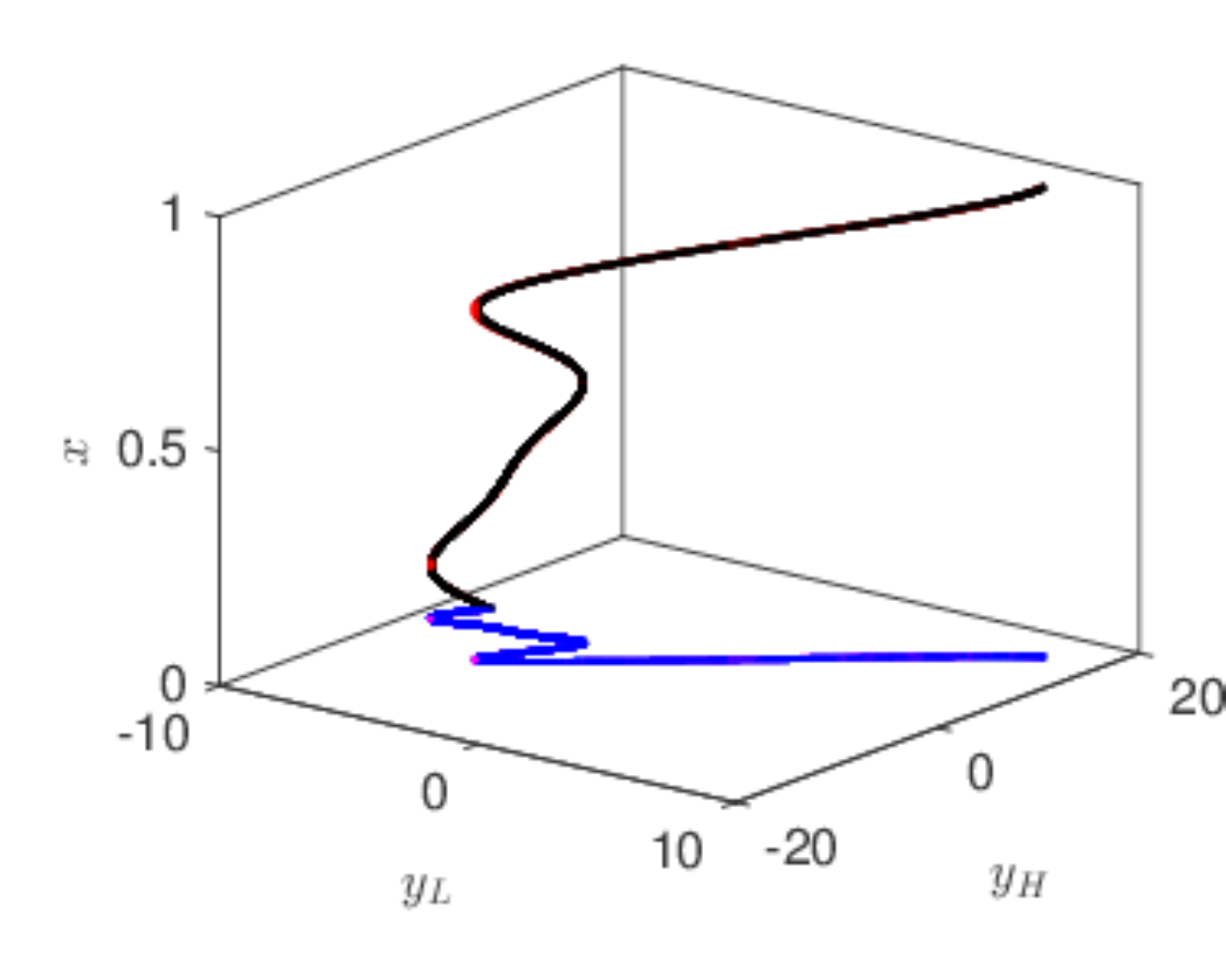}}
\caption{\label{caseI} Approximation of a continuous function from multi-fidelity data with {\it linear} correlation. 
(a) Training data at low- (11 data points) and high-fidelity levels (4 data points). 
(b) Predictions from DNN using high-fidelity data only; also included are the results of Kriging.
(c) Predictions from the multi-fidelity DNN (Red dashed line) and Co-Kriging \cite{forrester2007multi}) (Magenta dotted line).
(d) The Red dashed line in the $(x, y_L, y_H)$ plane represents Eq. \eqref{correlation} (on top of the exact Black solid line) and the Red dashed line in the $(y_L, y_H)$ plane represents the correlation discovered between the high- and low-fidelity data; the Blue solid line is the exact correlation.
}
\end{figure}

We first consider a pedagogical example of  approximating  an one-dimensional function based on data from two levels of fidelities.  The low-  and high-fidelity  data are generated from:
\begin{align}
y_L(x)  &= A(6x - 2)^2 \sin(12x - 4) + B(x - 0.5) + C,~ ~ x \in [0, 1]\\
y_H(x)  &= (6x - 2)^2 \sin(12x - 4), 
\end{align}
where $y_H$ is linear with $y_L$, and A = 0.5, B = 10, C = -5. As shown in Fig. \ref{caseIa}, the  training data at the  low- and high-fidelity level are  $x_L = \{ 0, 0.1, 0.2, 0.3, 0.4, 0.5, 0.6, \\ 0.7, 0.8, 0.9, 1\}$ and $x_{H} = \{ 0, 0.4, 0.6, 1\}$, respectively.  

We first try to predict the true function using the high-fidelity data only. For this case, we only need to keep $\mathcal{NN}_{H_2}$  (Fig. \ref{MPINNs}). In addition, the input for $\mathcal{NN}_{H_2}$ becomes $x$ because no low-fidelity data are available.  Here 4 hidden layers and 20 neurons per layer are adopted in  $\mathcal{NN}_{H_2}$ and no regularization is used. The learning rate is set as 0.001. As we can see in Fig. \ref{caseIb}, the present model provides inaccurate predictions due to the lack of sufficient high-fidelity data. Furthermore, we also plot the predictive posterior means of the Kriging \cite{forrester2007multi}, which is noted to be similar as the results from the $\mathcal{NN}_{H_2}$. 
Keeping the high-fidelity data fixed, we try to improve the accuracy of prediction by adding low-fidelity data (Fig. \ref{caseIa}).  In this case, the last DNN for the PDE is discarded.  Here 2 hidden layers and 20 neurons per layer are used in $\mathcal{NN}_L$, while 2 hidden layers with 10 neurons per layer are employed for $\mathcal{NN}_{H_2}$, and only 1 hidden with 10 neurons are used in $\mathcal{NN}_{H_1}$ (The size of $\mathcal{NN}_{H_1}$ is kept identical in all of the following cases).  The regularization rate is set to $\lambda = 10^{-2}$ with a learning rate 0.001. As shown in Fig. \ref{caseIc}, the present model provides accurate predictions for the high-fidelity profile.  In addition, the prediction using the Co-Kriging is displayed in Fig. \ref{caseIc} \cite{forrester2007multi}. We see that the learned profiles from these two methods are similar, while the result from the present model is slightly better than the Co-Kriging, which can be seen in the inset of Fig. \ref{caseIc}. Finally, the estimated  correlation is illustrated in Fig. \ref{caseId}, which also agrees quite well with the exact result.  Unlike the Co-Kriging/GPR, no prior knowledge on the correlation between the low- and high-fidelity data is needed in the multi-fidelity DNN, indicating that the present model can learn the correlation dynamically based on the given data.

The size of the neural network (e.g., depth and width) has a strong effect on the predictive accuracy \cite{raissi2019physics}, which is also investigated here. Since we have sufficient  low-fidelity data, it is easy to find an appropriate size for $\mathcal{NN}_L$ to approximate the low-fidelity function. Therefore, the particular focus is put on the size of  $\mathcal{NN}_{H_2}$ due to the fact that the few high-fidelity data  may yield overfitting.  Note that since the correlation between the low- and high-fidelity data is relatively simple, there is no need to set the $\mathcal{NN}_{H_2}$ to have a large size. Hence, we limit the ranges of the depth (i.e., $l$) and width (i.e., $w$) as: $l \in [1, 4]$ and $w \in [2, 32]$, respectively. Considering that a random initialization is utilized, we perform ten runs for each case with different depth and width. The mean and standard deviation for the relative $L_2$ errors defined as
\begin{align}\label{error}
E = \frac{1}{N}\sum^{n = N}_{n=1} \sqrt{\frac{\sum_j (y^*_j - y_j)^2}{\sum y^2_j}}, ~ \sigma =\sqrt{ \frac{\sum^{n = N}_{n=1}(E_n - E)^2}{N}},
\end{align}
are used to quantify the effect of the size of the $\mathcal{NN}_{H_2}$.  In Eq. \eqref{error}, $E$ is the mean relative $L_2$ errors, $n$ is the index of each run, $N$ is the total number of runs ($N = 10$), $j$ is the index for each sample data points, $E_n$ is the relative $L_2$ error for the $n-th$ run, and the definitions of $y^*$ and $y$ are the same as those in Sec. \ref{MPINNs}. As shown in Table \ref{sizeeffect}, the computational errors for $\mathcal{NN}_{H_2}$ with different depth and width are almost the same. In addition, the standard deviation for the relative $L_2$ errors are not presented because they are less than $10^{-5}$ for each case.  
All these results demonstrate the robustness of the 
multi-fidelity DNNs.  To reduce the computational cost as well and retain the accuracy, a good choice for the size of $\mathcal{NN}_{H_2}$  may be $l \in [1,2]$ and $w \in [4, 20]$ in low dimensions.

\begin{table}[htbp]
\centering
 \caption{\label{sizeeffect}Mean relative $L_2 (\times 10^{-3})$ for NNs with different sizes.}
 \begin{tabular}{c|cccc}
  \hline \hline
  \diagbox{Depth}{Width} & 4 & 8 &16 &32\\ \hline
   1 & 3.1 & 3.0 & 4.6  &2.9 \\ \hline
   2& 3.4 & 3.0 & 3.0  &3.1 \\ \hline
   3 & 3.1 & 3.1 & 3.1  &3.0\\ \hline
   4 & 3.0 & 3.0 & 3.0  &3.0 \\
  \hline \hline
 \end{tabular}
\end{table}

\subsubsection{Discontinuous function with linear correlation}
As mentioned in \cite{raissi2016deep}, the approximation of a discontinuous function using GPR is challenging due to the continuous kernel employed.  We then proceed to test the  capability of the present model for approximating discontinuous functions. The low- and high-fidelity data are generated by the following ``Forrester" functions with jump \cite{raissi2016deep}: 
\begin{equation}
y_L(x) =\left \{
\begin{aligned}
&0.5(6x - 2)^2\sin(12x - 4) + 10(x - 0.5) - 5, ~ 0 \le x \le 0.5,\\
&3 + 0.5(6x - 2)^2\sin(12x - 4) + 10(x - 0.5) - 5, ~ 0.5 < x \le 1,
\end{aligned}
\right.
\end{equation}
and
\begin{equation}
y_H(x) =\left \{
\begin{aligned}
&2y_L(x) - 20x + 20, ~ 0 \le x \le 0.5,\\
&4 + 2y_L(x) - 20x + 20, ~ 0.5 < x \le 1.
\end{aligned}
\right.
\end{equation}

\begin{figure}
\centering
\subfigure[]{\label{caseIIa}
\includegraphics[width = 0.45\textwidth]{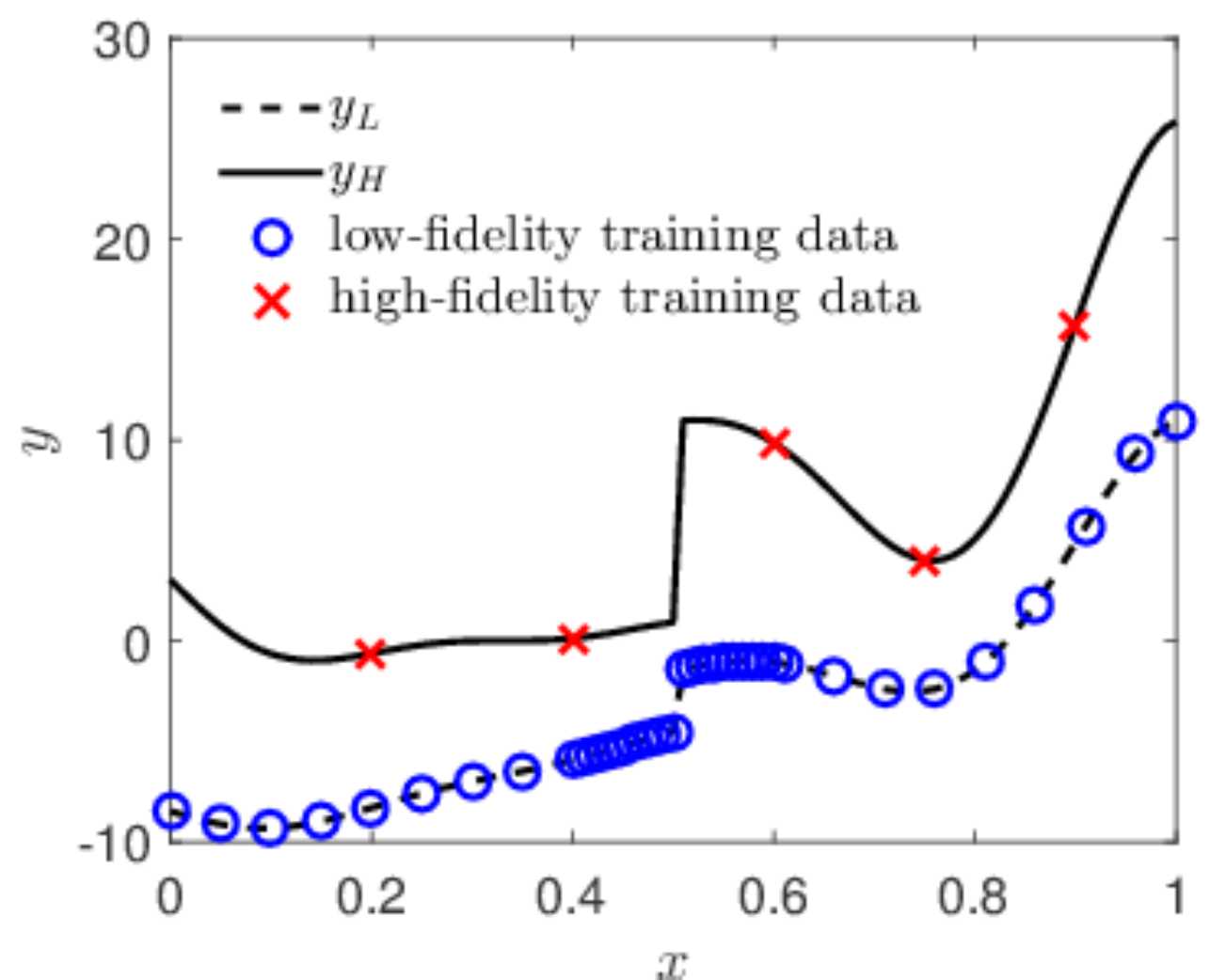}}
\subfigure[]{\label{caseIIb}
\includegraphics[width = 0.45\textwidth]{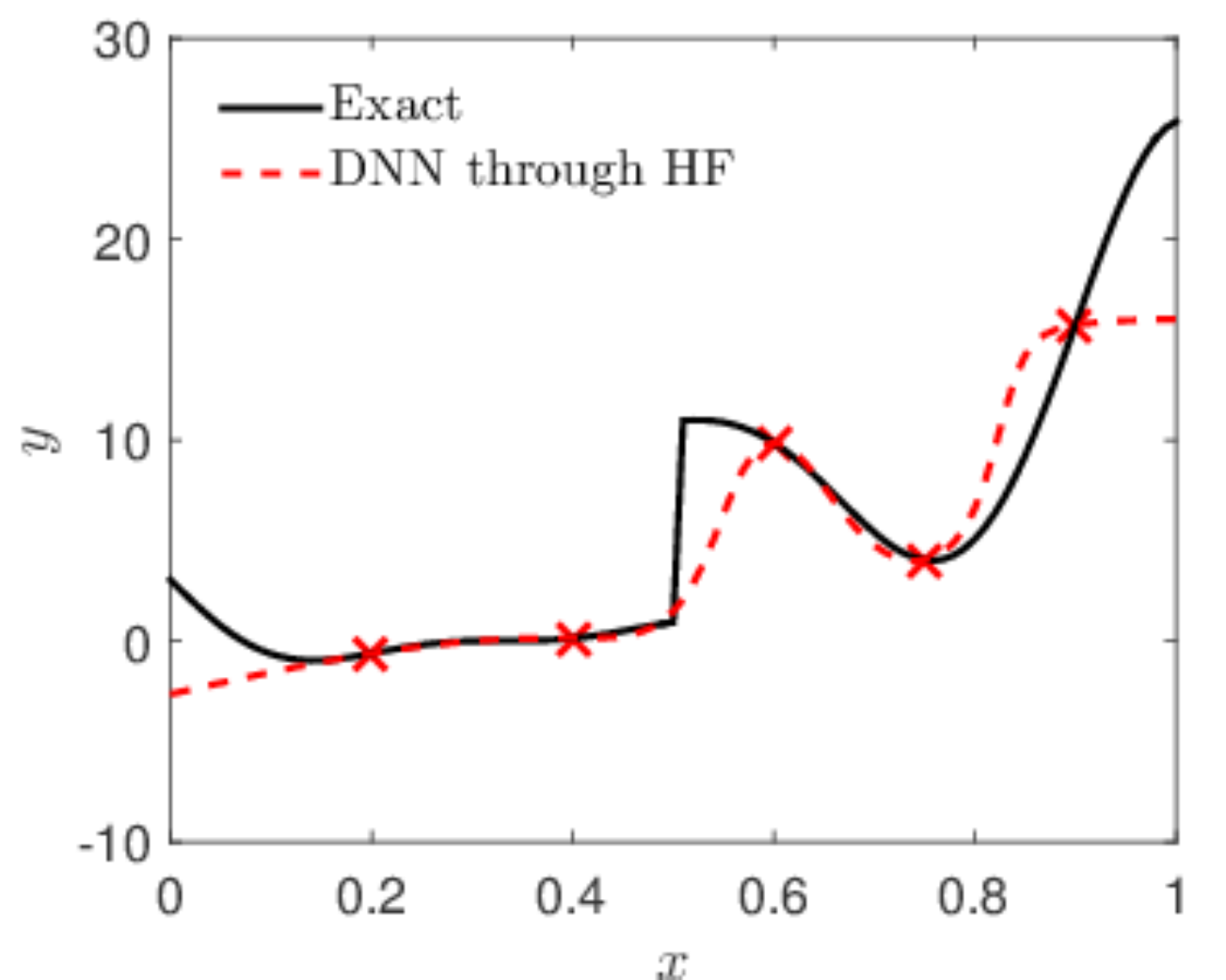}}
\subfigure[]{\label{caseIIc}
\includegraphics[width = 0.45\textwidth]{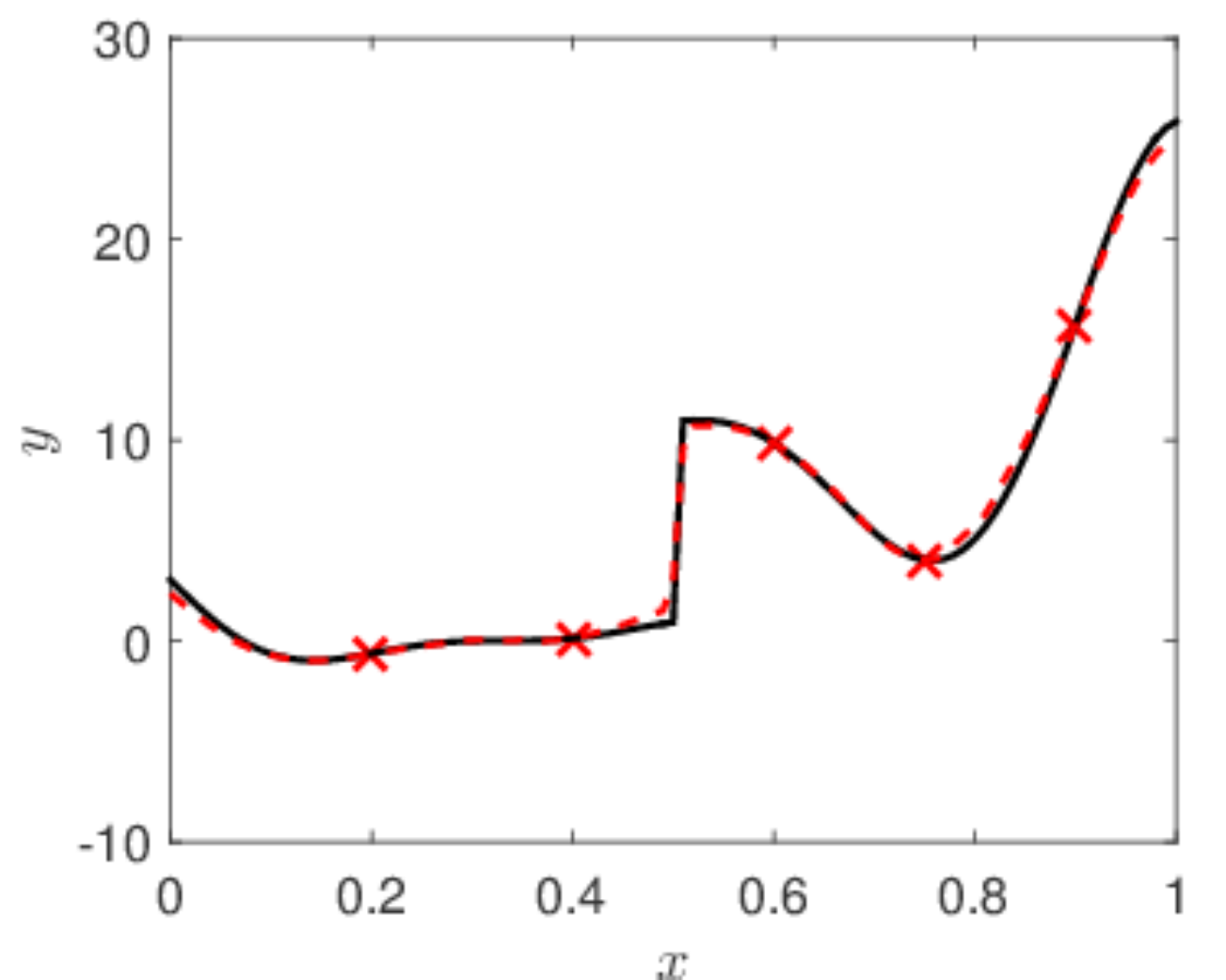}}
\subfigure[]{\label{caseIId}
\includegraphics[width = 0.45\textwidth]{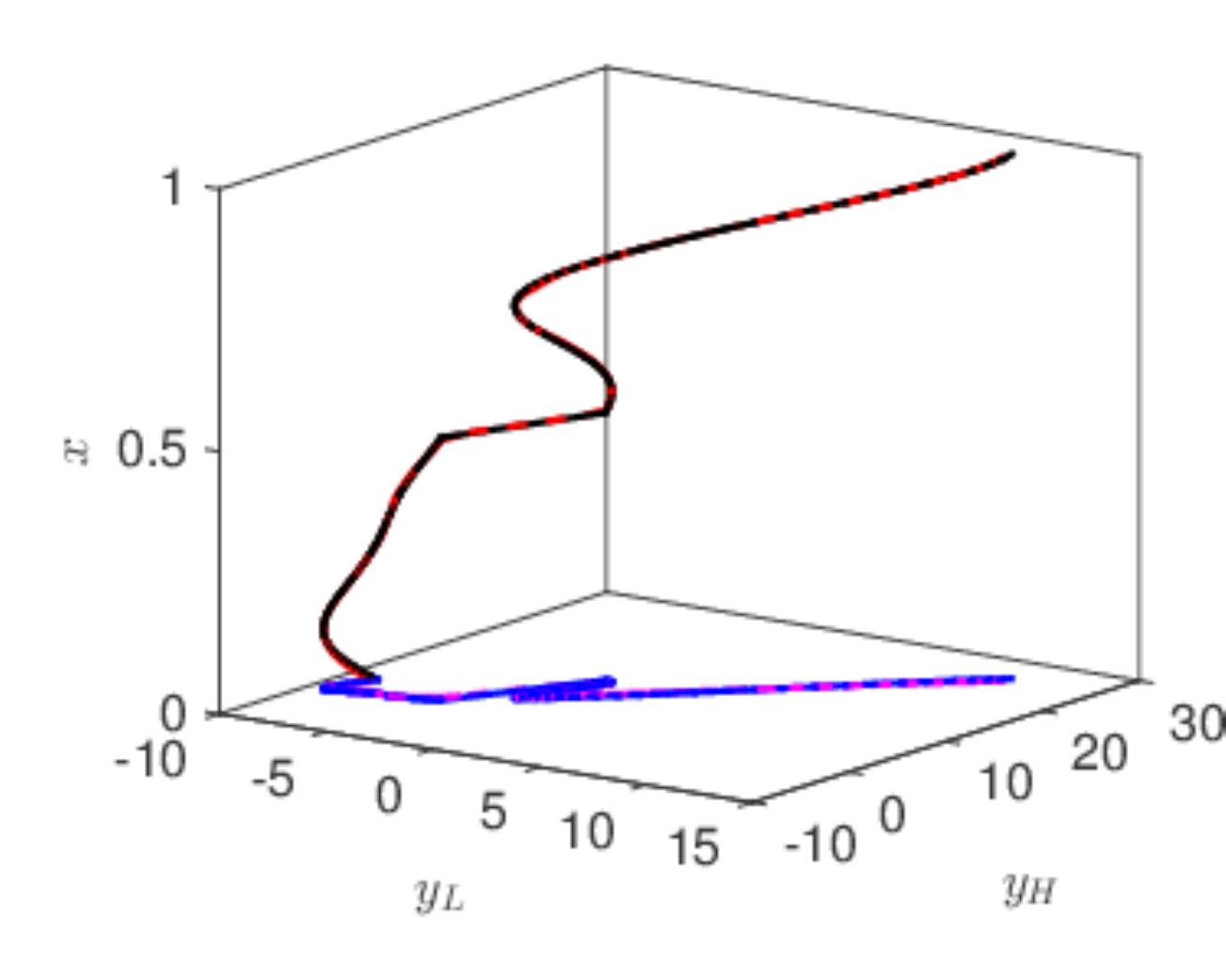}}
\caption{\label{caseII} Approximation of a discontinuous function from multi-fidelity data with {\it linear} correlation. 
(a) Training data at low- (38 data points) and high-fidelity levels (5 data points).
(b) Predictions from DNN using high-fidelity data only (Red dash line); also included is the exact curve (Black solid line). 
(c) Predictions from multi-fidelity DNN for high-fidelity (Red dash line).
(d) The Red dashed line in the $(x, y_L, y_H)$ plane represents Eq. \eqref{correlation} (on top of the exact Black solid line) and the Red dashed line in the $(y_L, y_H)$ plane represents the
 correlation discovered between the high- and low-fidelity data; the Blue line is the exact correlation.
}
\end{figure}

As illustrated in Fig. \ref{caseIIa},  38 and 5 sampling data points are employed as the training data at the low- and high-fidelity level, respectively. The learning rate is again set as 0.001 for all test cases here. Similarly, we employ the $\mathcal{NN}_{H_2}$ ($l \times w = 4 \times 20$) to predict the high-fidelity values on the basis of the given high-fidelity data only, but
the corresponding prediction is not good (Fig. \ref{caseIIb}). However, using the multi-fidelity data, the present model can provide quite accurate predictions for the high-fidelity profile (Fig. \ref{caseIIc}).  Remarkably, the multi-fidelity DNN can capture the discontinuity at $x = 0.5$ at the high-fidelity level quite well even though no data are available in the range $0.4 < x < 0.6$. This is reasonable because the low- and high-fidelity data share the same trend as $0.4< x < 0.6$, yielding the correct predictions of the high-fidelity values in this zone.  Furthermore, the learned correlation is displayed in Fig.  \ref{caseIId}, which shows only slight differences from the exact correlation.

\subsubsection{Continuous function with nonlinear correlation}
\begin{figure}
\centering
\subfigure[]{\label{caseIIIa}
\includegraphics[width = 0.45\textwidth]{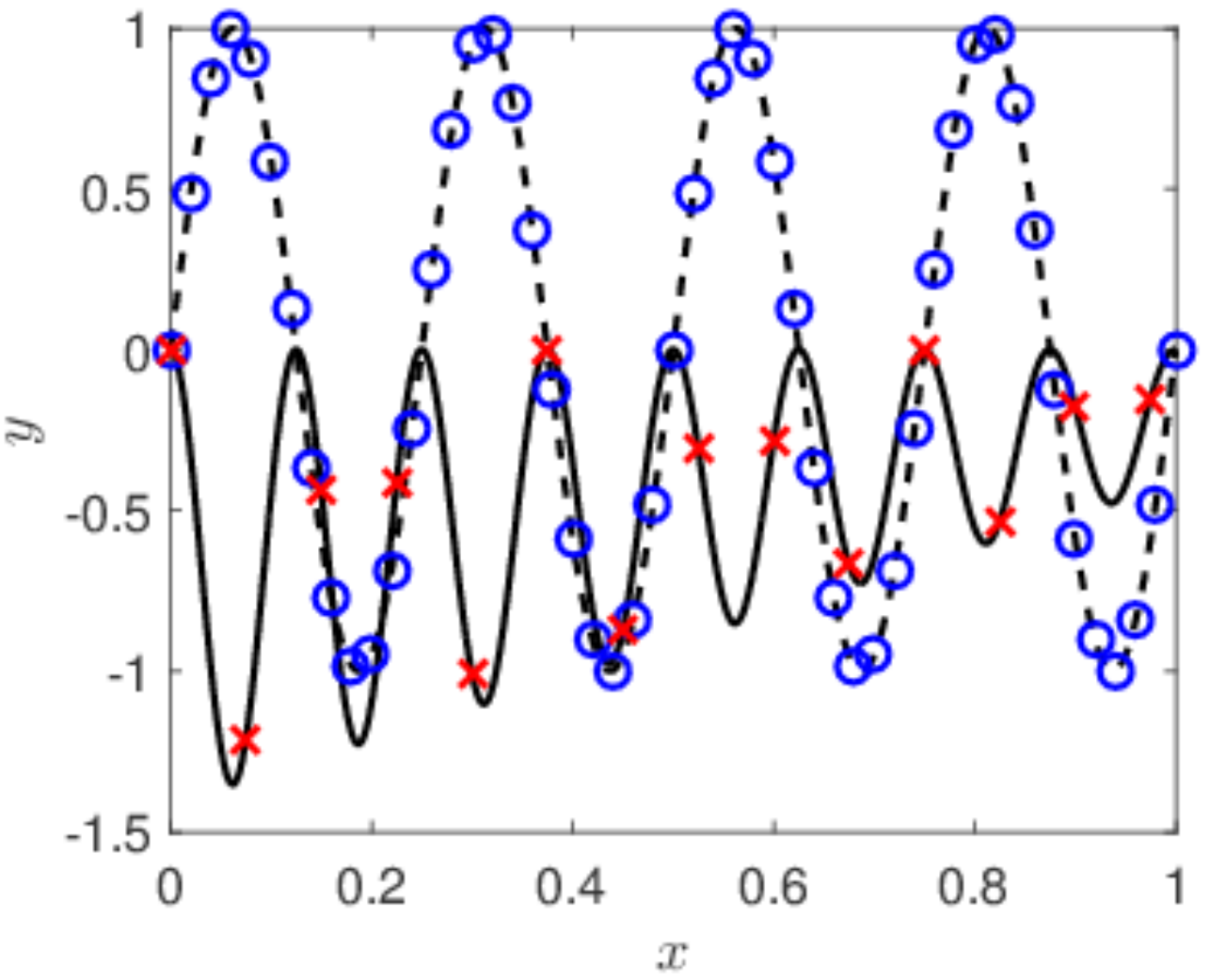}}
\subfigure[]{\label{caseIIIb}
\includegraphics[width = 0.45\textwidth]{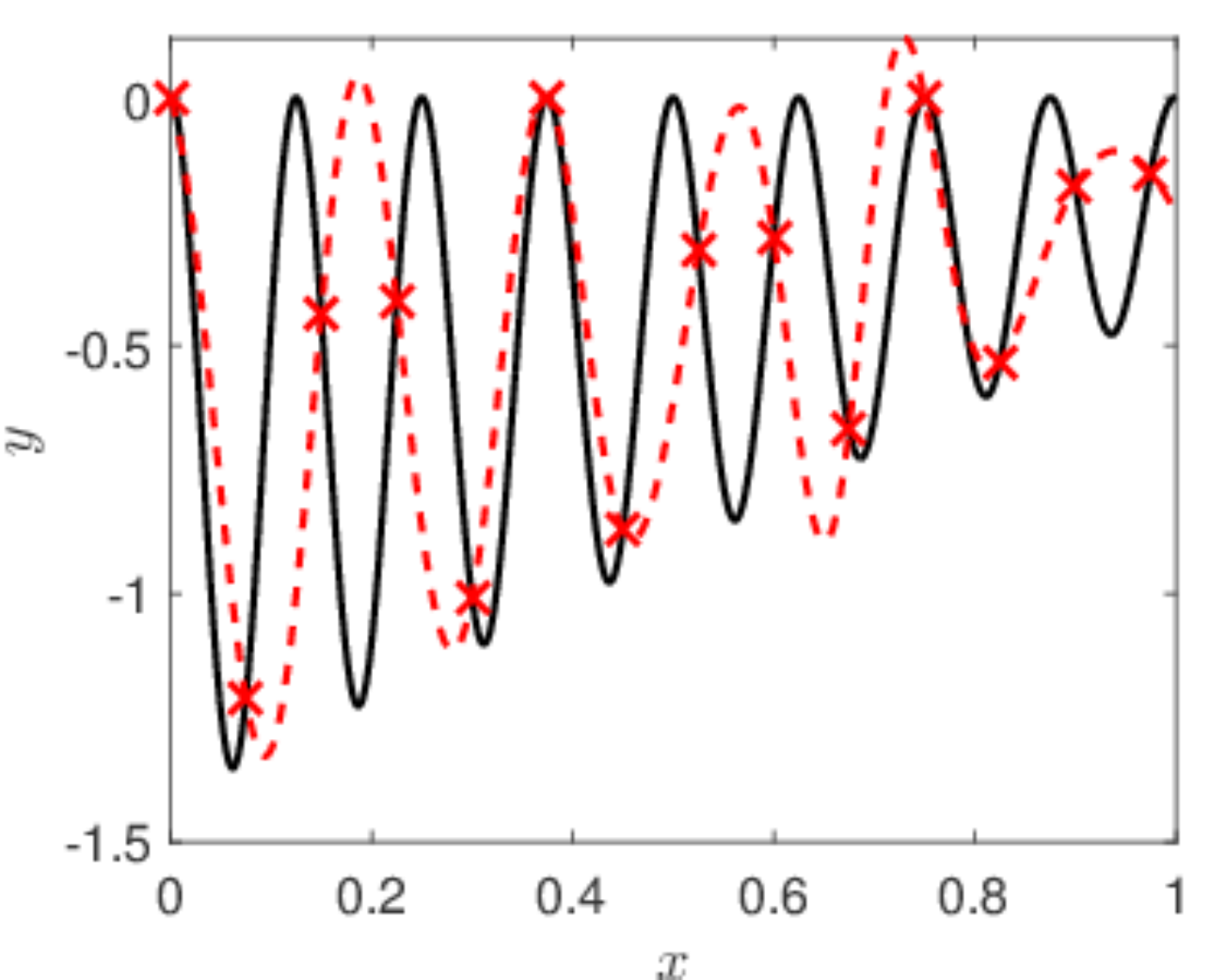}}
\subfigure[]{\label{caseIIIc}
\includegraphics[width = 0.45\textwidth]{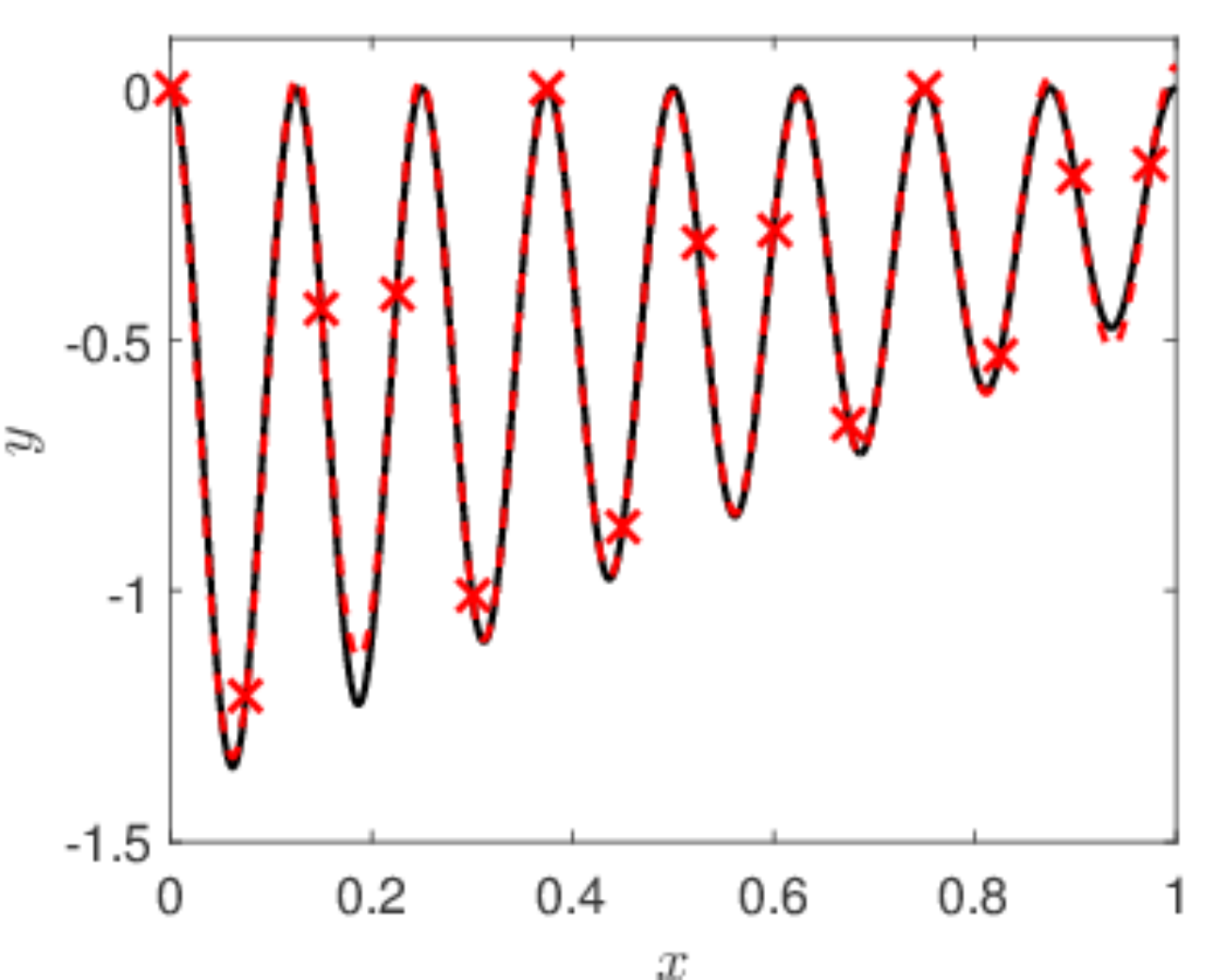}}
\subfigure[]{\label{caseIIId}
\includegraphics[width = 0.45\textwidth]{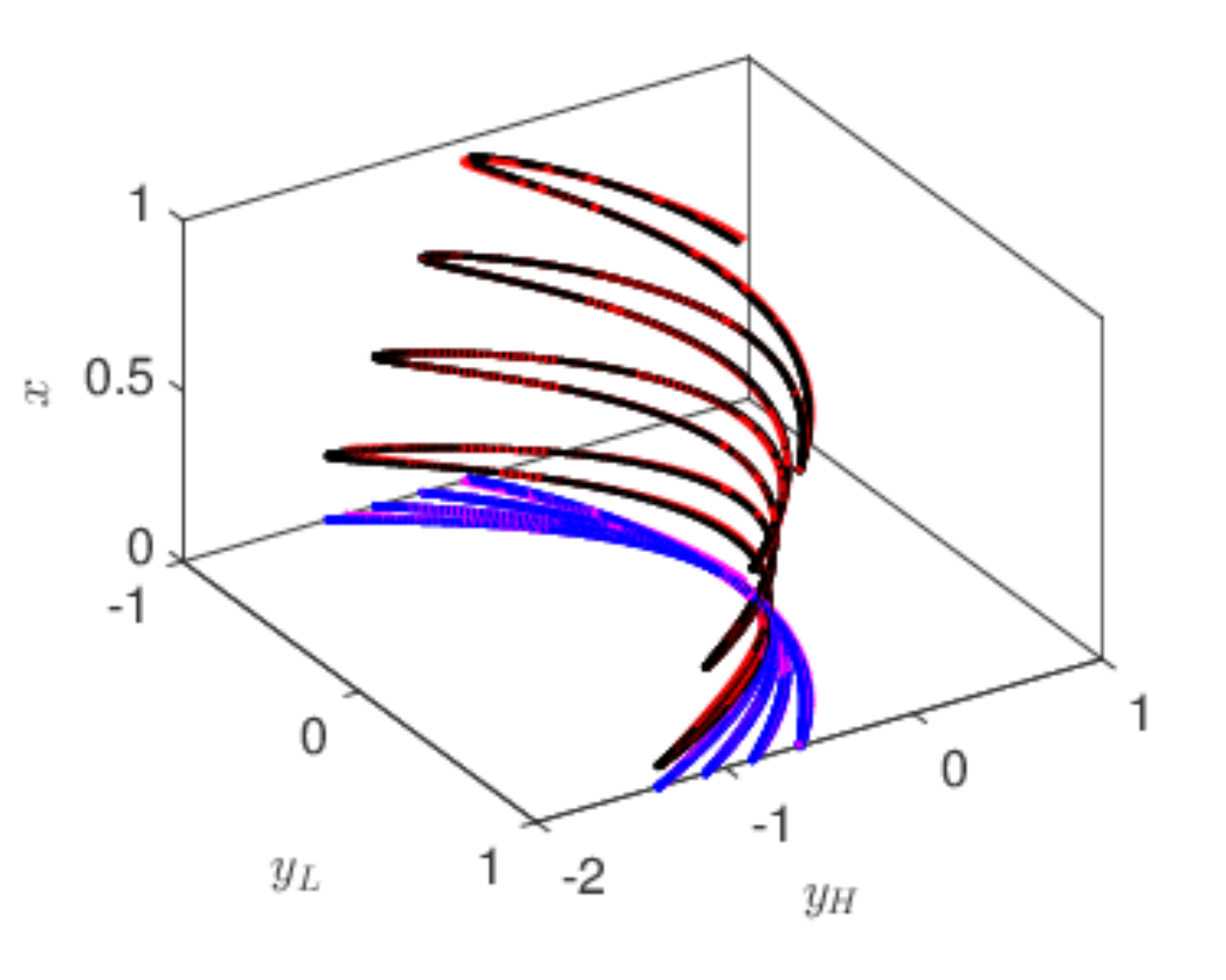}}
\caption{\label{caseIII}
Approximation of a continuous function from multi-fidelity data with {\it nonlinear} correlation. 
(a) Training data at low- (51 data points) and high-fidelity levels (14 data points). Black solid line: High-fidelity values, Black dashed line: Low-fidelity values, Red cross: High-fidelity training data, Blue circle: Low-fidelity training data. 
(b) Predictions from high-fidelity DNN (Red dashed line); Black solid line: Exact values. 
(c) Predictions from multi-fidelity DNN for high-fidelity (Red dash line).
(d) The Red dashed line in $(x,y_L,y_H)$ represents Eq. \eqref{correlation} (on top of the exact Black solid line) and the Red dashed line in
$(y_L,y_H)$ represents the correlation discovered between the high- and low-fidelity data; the Blue line is the exact correlation.
 }
\end{figure}

To test the present model for capturing complicated nonlinear correlations between the low- and high-fidelity data, we further consider the following case \cite{perdikaris2017nonlinear}:
\begin{align}
y_L(x) &= \sin(8 \pi x), ~x \in [0, 1], \\
y_H(x) &= (x - \sqrt{2}) y^2_L(x).
\end{align}
Here, we employ 51 and 14 data points (uniformly distributed) for low- and high-fidelity, respectively,  as the training data, (Fig. \ref{caseIIIa}). The learning rate for all test cases is still 0.001.
As before, the $\mathcal{NN}_{H_2}$ ($l \times w = 4 \times 20$) cannot provide accurate predictions for the high-fidelity values using only the few high-fidelity data points as displayed in Fig. \ref{caseIIIb}. We then test the performance of the multi-fidelity DNN based on the multi-fidelity training data. Four hidden layers and 20 neurons per layer are used in $\mathcal{NN}_L$, and 2 hidden layers with 10 neurons per layer are utilized for $\mathcal{NN}_{H_2}$.  Again, the predicted profile from the present model agrees well with the exact profile at the  high-fidelity level, as shown in Fig. \ref{caseIIIc}. It is interesting to find that the multi-fidelity DNN can still provide accurate predictions for the high-fidelity profile even though the trend for the low-fidelity data is opposite to that of the high-fidelity data, e.g., $0 < x < 0.2$, a case of {\it adversarial} type of data. In addition, the learned correlation between the low- and high-fidelity data agree well with the exact one as illustrated in Fig. \ref{caseIIId}, indicating that the multi-fidelity DNN is capable of  discovering the non-trivial underlying correlation on the basis of training data.

\subsubsection{Phase-shifted oscillations}
\begin{figure}
\centering
\subfigure[]{\label{phaseinitiala}
\includegraphics[width = 0.45\textwidth]{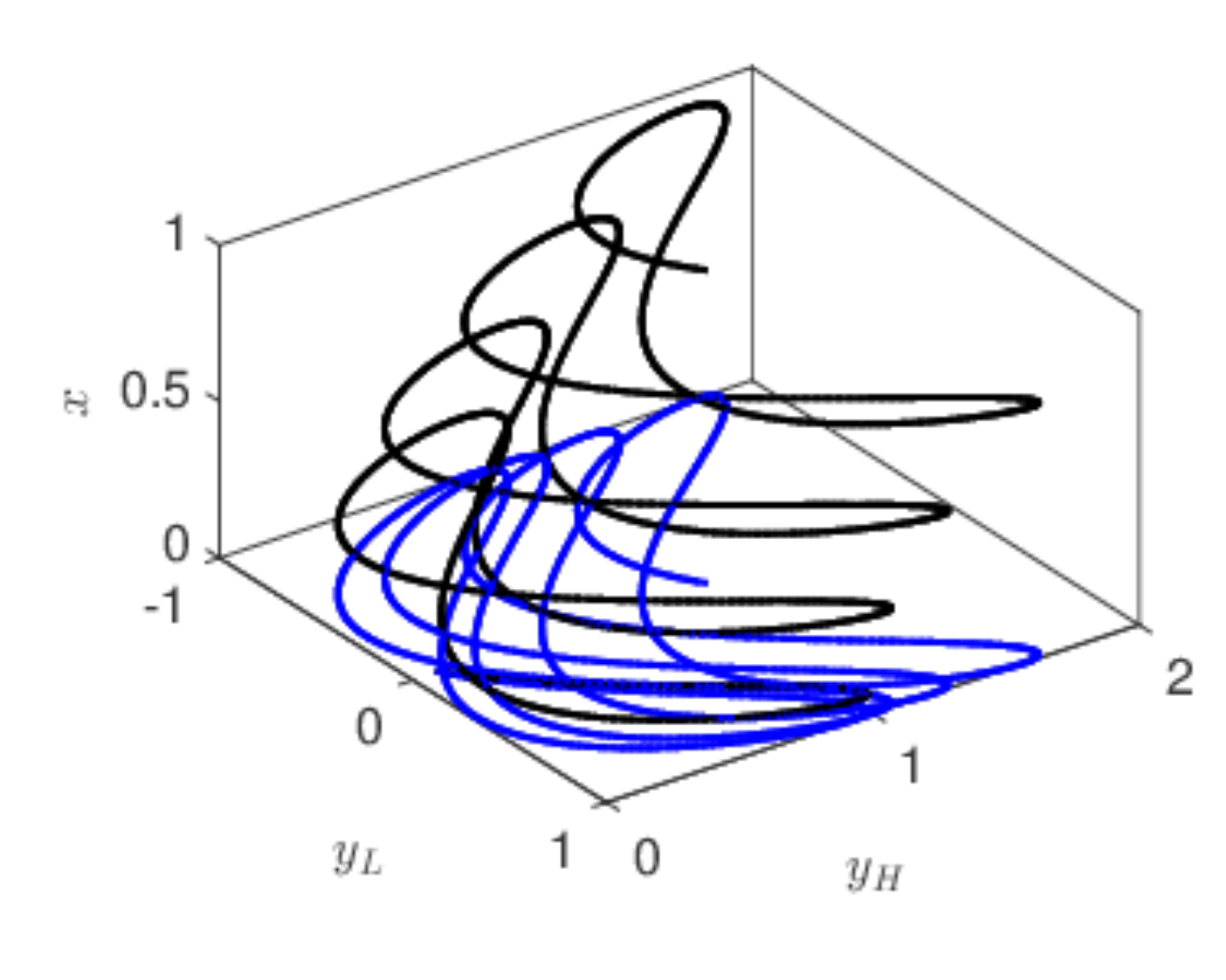}}
\subfigure[]{\label{phaseinitialb}
\includegraphics[width = 0.45\textwidth]{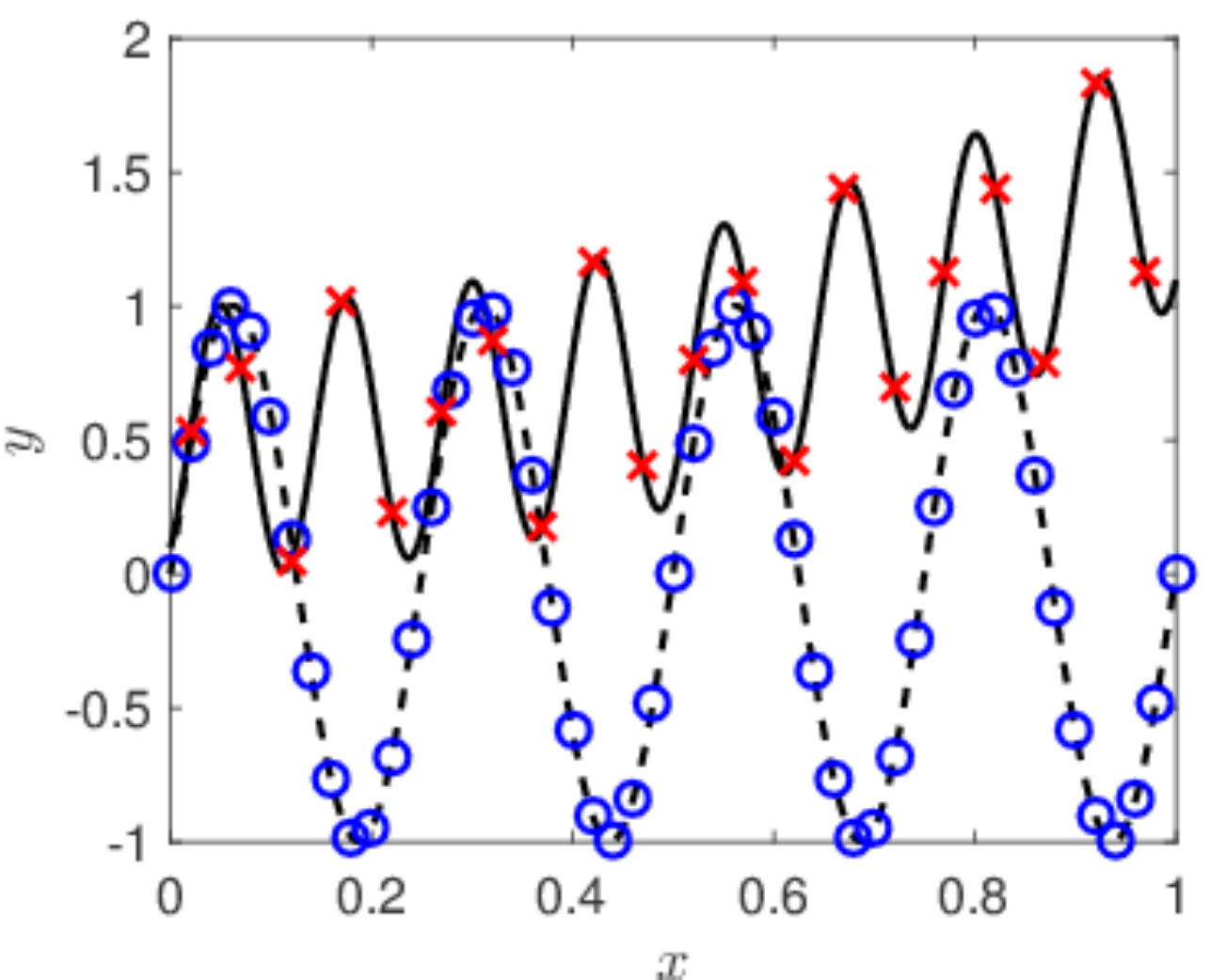}}
\subfigure[]{\label{phasea}
\includegraphics[width = 0.45\textwidth]{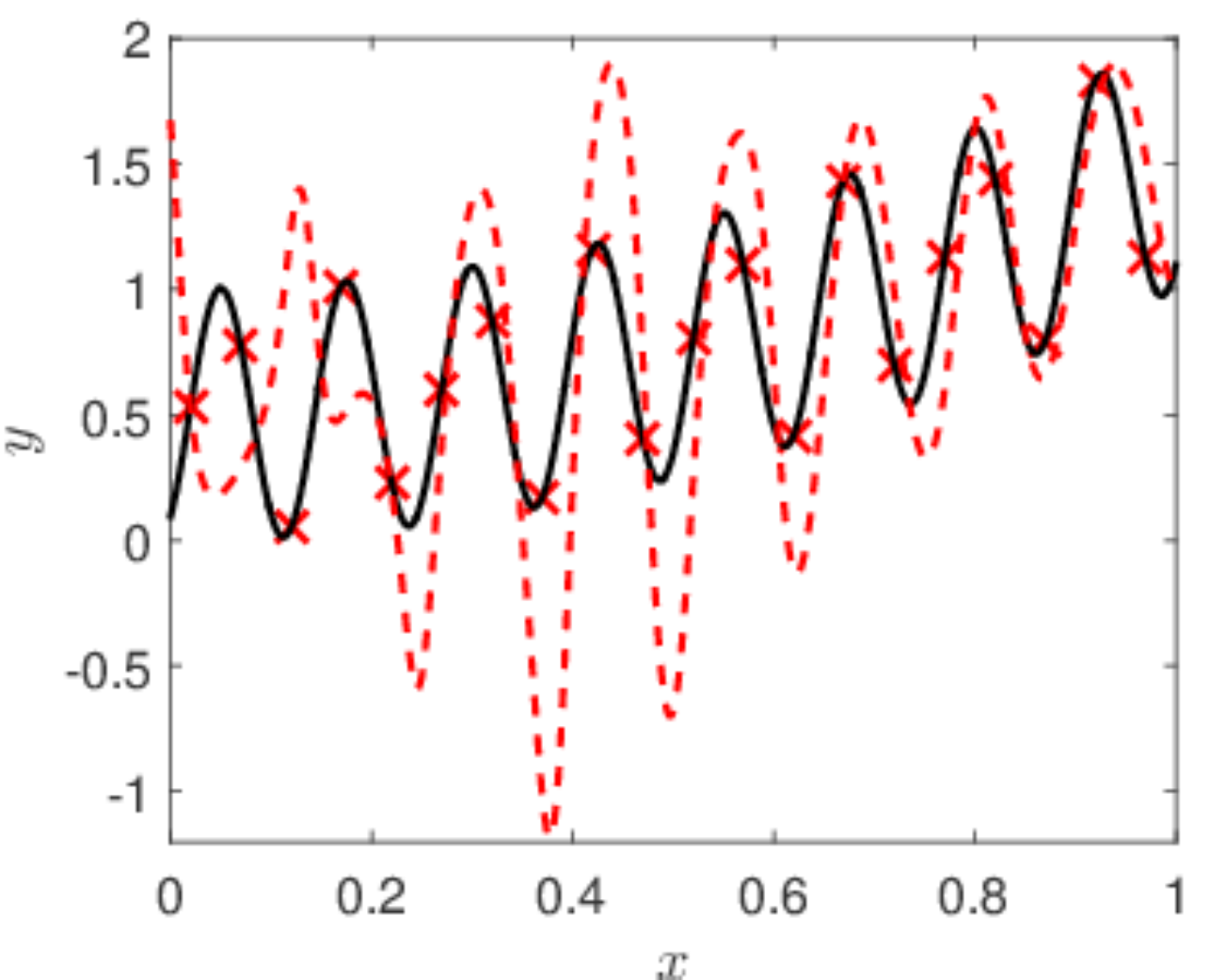}}
\subfigure[]{\label{phaseb}
\includegraphics[width = 0.45\textwidth]{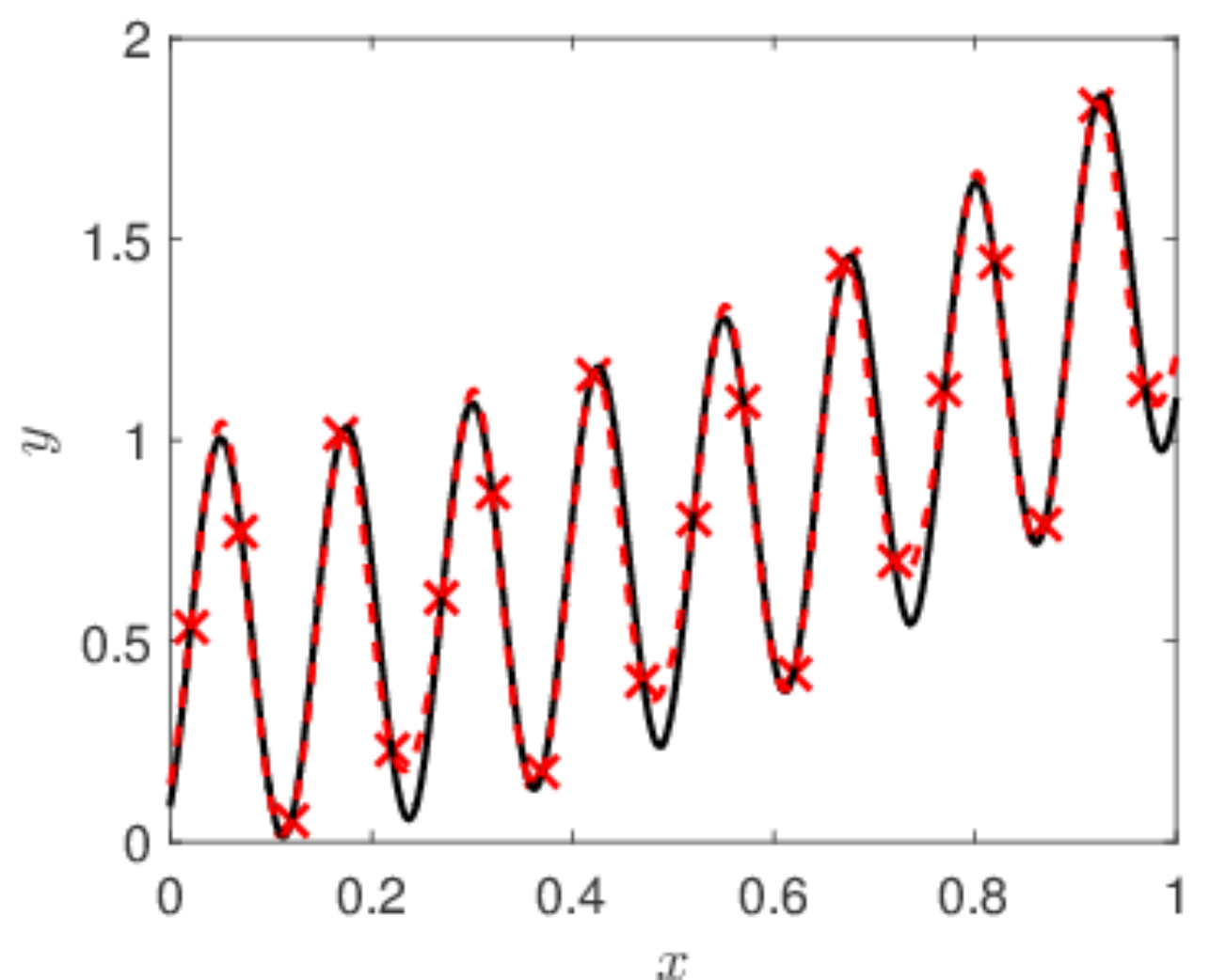}}
\caption{\label{phaseinitial}
Approximation of continuous function from multi-fidelity data with phase-shifted oscillations and {\it highly-nonlinear} correlation: 
(a) Correlation among $x$, $y_L$, and $y_H$. The Blue line represents the 
projection in the $(y_L,y_H)$ plane. 
(b) Training data for $y_L$ and $y_H$. Black solid line: Exact high-fidelity values; Black dashed line: Exact low-fidelity values; Red cross: High-fidelity training data; Blue circle: Low-fidelity training data.
(c) Predictions from Method I (without time-delay) (Red dashed line). 
(d) Predictions from Method II (with time-delay) (Red dashed line). The learned optimal value for $\tau$ is $4.49 \times 10^{-2}$.
}
\end{figure}

For more complicated correlations between the low and high-fidelity data, we can easily extend the 
multi-fidelity DNN based on the ``embedding theory" to enhance the capability for learning more complex correlations \cite{lee2018linking} (For more details on the embedding theory, refer to \ref{embedding}).  Here, we consider the following low-/high-fidelity functions with phase errors \cite{lee2018linking}:
\begin{align}\label{phase1}
y_H(x) &= x^2 + \sin^2(8 \pi x + \pi/10), \\
y_L(x) & = \sin(8 \pi x).
\end{align}
We can further write $y_H$ in terms of $y_L$ as
\begin{align}\label{phase2}
y_H = x^2 + (y_L \cos(\pi/10) + y^{(1)}_L \sin(\pi/10))^2,
\end{align}
where $y^{(1)}_L$ denotes the first derivatives of $y_L$.  The relation between the low- and high-fidelity data is displayed in Fig. \ref{phaseinitiala}, which is rather complicated. The performance of the 
multi-fidelity DNN for this case will be tested next.
To approximate the high-fidelity function, we select 51 and 16 uniformly distributed sample points as the training data for low- and high-fidelity values, respectively (Fig. \ref{phaseinitialb}). The selected learning rate for all test cases is 0.001. Here, we test two types of inputs for $\mathcal{NN}_{H_2}$, i.e., $[x, y_L(x)]$ (Method I), and $[x, y_L(x), y_L(x - \tau)]$ (Method II) ($\tau$ is the delay).  Four hidden layers and 20 neurons per layer are used in $\mathcal{NN}_L$, and 2 hidden layers with 10 neurons per layer are utilized for $\mathcal{NN}_{H_2}$.  As shown in Fig. \ref{phaseinitial}, it is interesting to find that Method II provides accurate predictions for the high-fidelity values (Fig. \ref{phaseb}), while  Method I fails (Fig. \ref{phasea}).  As mentioned in \cite{lee2018linking}, the term $y_L(x - \tau)$ can be viewed as an implicit approximation for $y^{(1)}_L$, which enables Method II to capture the correlation in Eq. \eqref{phase2} based only on a small number of high-fidelity data points. However, given that no information on $y^{(1)}_L$ is available in Method I, the present datasets are insufficient to obtain the correct correlation. 

\subsubsection{20-dimensional function approximation}
\begin{figure}
\centering
\subfigure[]{\label{hdpreda}
\includegraphics[width = 0.45\textwidth]{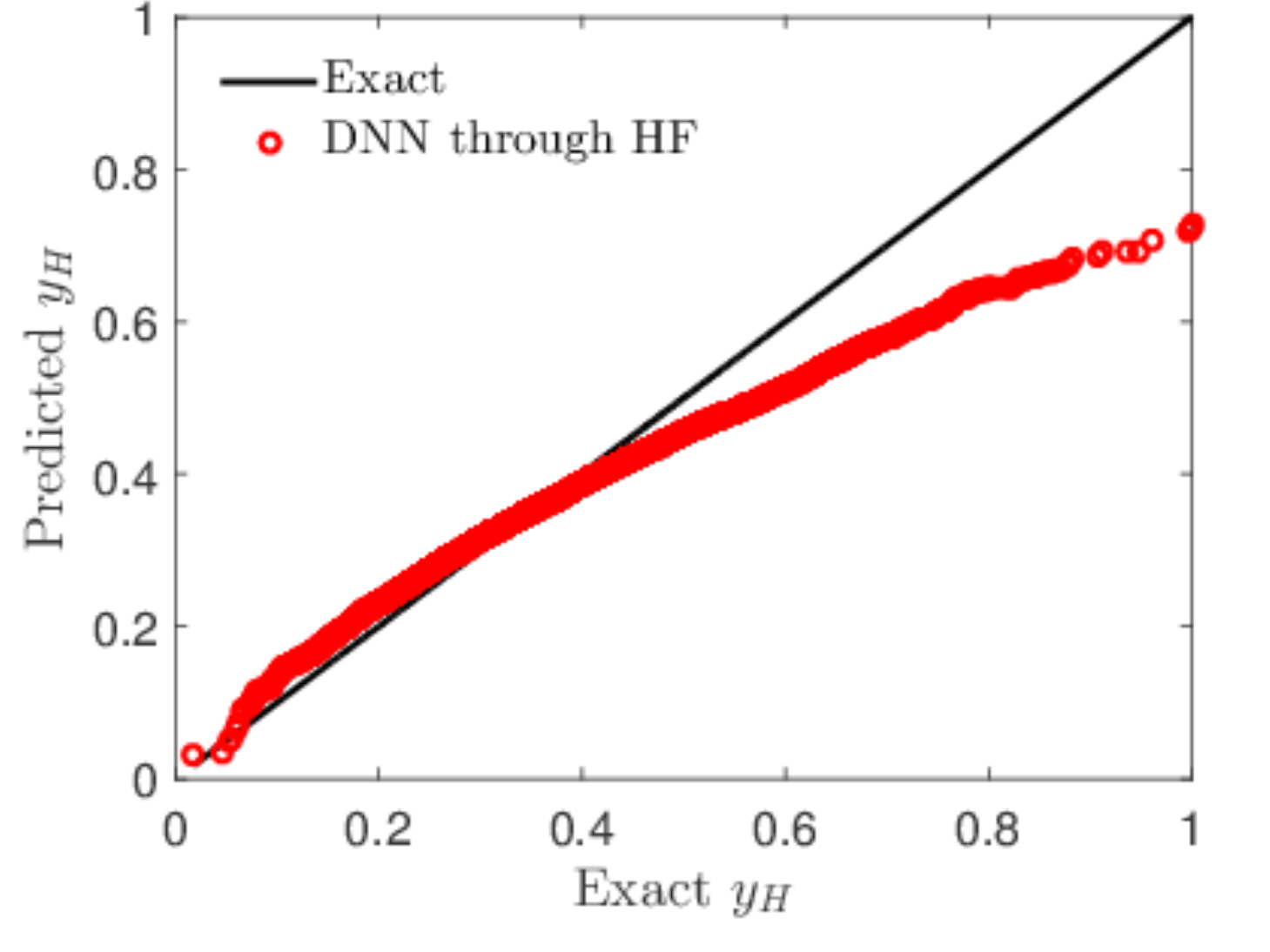}}
\subfigure[]{\label{hdpredb}
\includegraphics[width = 0.45\textwidth]{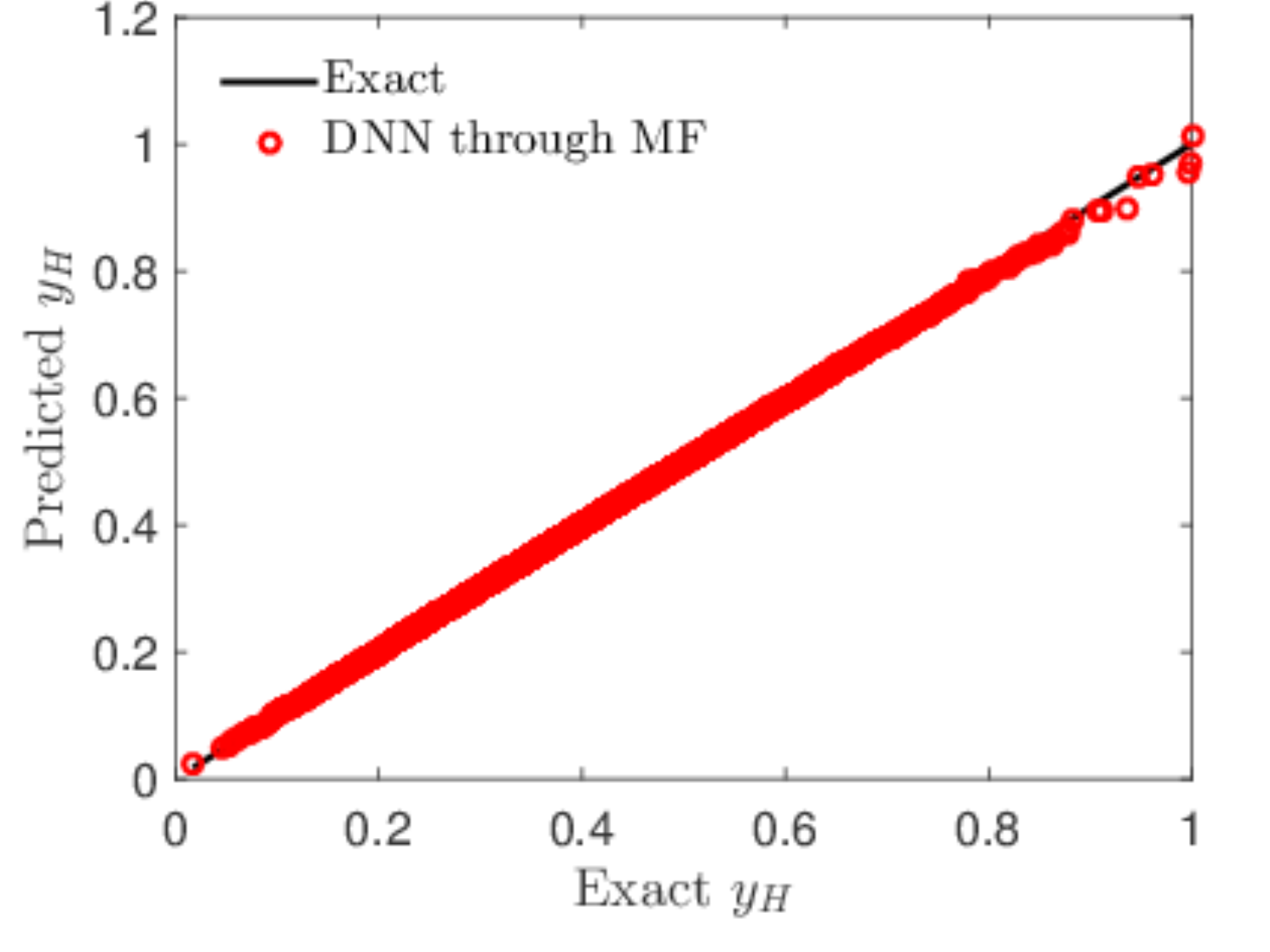}}
\caption{\label{hdpred} 
Approximations of the 20-dimensional function (learning rate: 0.001). 
(a) Single-fidelity predictions from  high-fidelity data. $\mathcal{NN}_{H_2} \rightarrow 4 \times 160$ with 5000 randomly selected high-fidelity data, and 10000 test data at random locations. 
(b) Multi-fidelity DNN predictions. $\mathcal{NN}_L \rightarrow 4 \times 128$, $\mathcal{NN}_{H_2} \rightarrow 2 \times 64$ with 30000 and 5000 randomly selected low-/high-fidelity data, and 10000 test data at random locations.}
\end{figure}

In principle, the new multi-fidelity DNN can approximate any high-dimensional function so here we 
take a modest size so that is not computationally expensive to train the DNN. Specifically, we generate 
the low- and high-fidelity data for a 20-dimensional function from the following equations: \cite{shan2010metamodeling}
\begin{align}
    y_H(x) &= (x_1 - 1)^2 + \sum^{20}_{i=2} \left(2x^2_i - x_{i-1}\right)^2, x_i \in[-3, 3], i = 1, 2, ..., 20, \\
    y_L(x) &= 0.8y_H(x) - \sum^{19}_{i=1} 0.4 x_i x_{i+1} - 50.
\end{align}
As shown in Fig. \ref{hdpreda}, using only the available high-fidelity data does not lead to an accurate
function approximation but using the multi-fidelity DNN approach gives excellent results as shown in 
Fig. \ref{hdpredb}.

\vspace{0.05in}
In summary, in this section we have demonstrated using different data sets and correlations that multi-fidelity DNNs can adaptively learn the underlying correlation between the low- and high-fidelity data from the given datasets without any prior assumption on the correlation. In addition, they can be applied to high-dimensional cases, hence outperforming GPR \cite{perdikaris2017nonlinear}. Finally, the present framework can be easily extended based on the embedding theory to non-functional correlations, which enables multi-fidelity DNNs to learn more complicated nonlinear correlations induced by phase errors of the low-fidelity data (adversarial data).

\subsection{Inverse PDE problems with nonlinearities}
In this section, we will apply the multi-fidelity PINNs (MPINNs) to two inverse PDE problems with nonlinearities, specifically, unsaturated flows and reactive transport in porous media. 
We assume that the hydraulic conductivity is first estimated based on scarce high-fidelity measurements of the pressure head. Subsequently, the reactive models are further learned given a small set of  high-fidelity observations of the solute concentration.

\subsubsection{Learning the hydraulic conductivity for nonlinear unsaturated flows}


Unsaturated flows play an important role in the ground-subsurface water interaction zone \cite{markstrom2008gsflow,hayashi2002effects}.  Here we consider a steady unsaturated flow in an one-dimensional (1D) column with a variable water content,  which can be described by the following equation as
\begin{equation}
\begin{aligned}\label{flows}
\partial_x \left( K(h) \partial_x h \right) = 0.
\end{aligned}
\end{equation}
We consider two types of boundary conditions, i.e., (1)  constant flux at the inlet and constant pressure head at the outlet, $q = -K\partial_x h = q_0, ~x = 0; ~h = h_1, ~x = L_x$ (Case I),  and (2) constant pressure head at both the inlet and outlet, $h = h_0, ~x = 0; ~h = h_1, ~x = L_x$ (Case II). Here $L_x = 200cm$ is the length of the column, $h$ is the pressure head, $h_0$ and $h_1$ are, respectively, the pressure head at the inlet and outlet, $q$ represents the flux, and $q_0$ is the flux at the inlet, which is a constant. In addition, $K(h)$ denotes the pressure-dependent hydraulic conductivity, which is expressed as  
\begin{align}\label{per}
K(h) = K_s S^{1/2}_e \left[ 1 - (1 - S^{1/m}_e)^m\right]^{2},
\end{align}
where $K_s$ is the saturated hydraulic conductivity, and $S_e$ is the effective saturation that is a function of $h$. It is noted that several models have been developed to characterize $S_e$ but among them, the van Genuchten model is the most widely used \cite{van1980closed}, which reads as follows:
\begin{align}\label{vgp}
S_e = \frac{1}{(1 + |\alpha_0 h|^n)^{m}},  ~ m = 1 - 1/n.
\end{align}
In Eq. \eqref{vgp}, $\alpha_0$ is related to the inverse of the air entry suction, and $m$ represents a measure of the pore-size distribution. To obtain the velocity field for later applications, we should first obtain the distribution of $K(h)$. Unfortunately, both parameters depend on the geometry of porous medium and are difficult to measure directly. We note that the pressure head can be measured more easily in comparison to $\alpha_0$ and $m$. Therefore, we assume that partial measurements of $h$ are available without the direct measurements of $\alpha_0$ and $m$. The objective is to estimate $\alpha_0$ and $m$ based on the observations of $h$. Then, we can compute the distribution of $K(h)$ according to Eqs. \eqref{per} and \eqref{vgp}.

\begin{figure}
\centering
\subfigure[]{\label{difflhflowa}
\includegraphics[width = 0.45\textwidth]{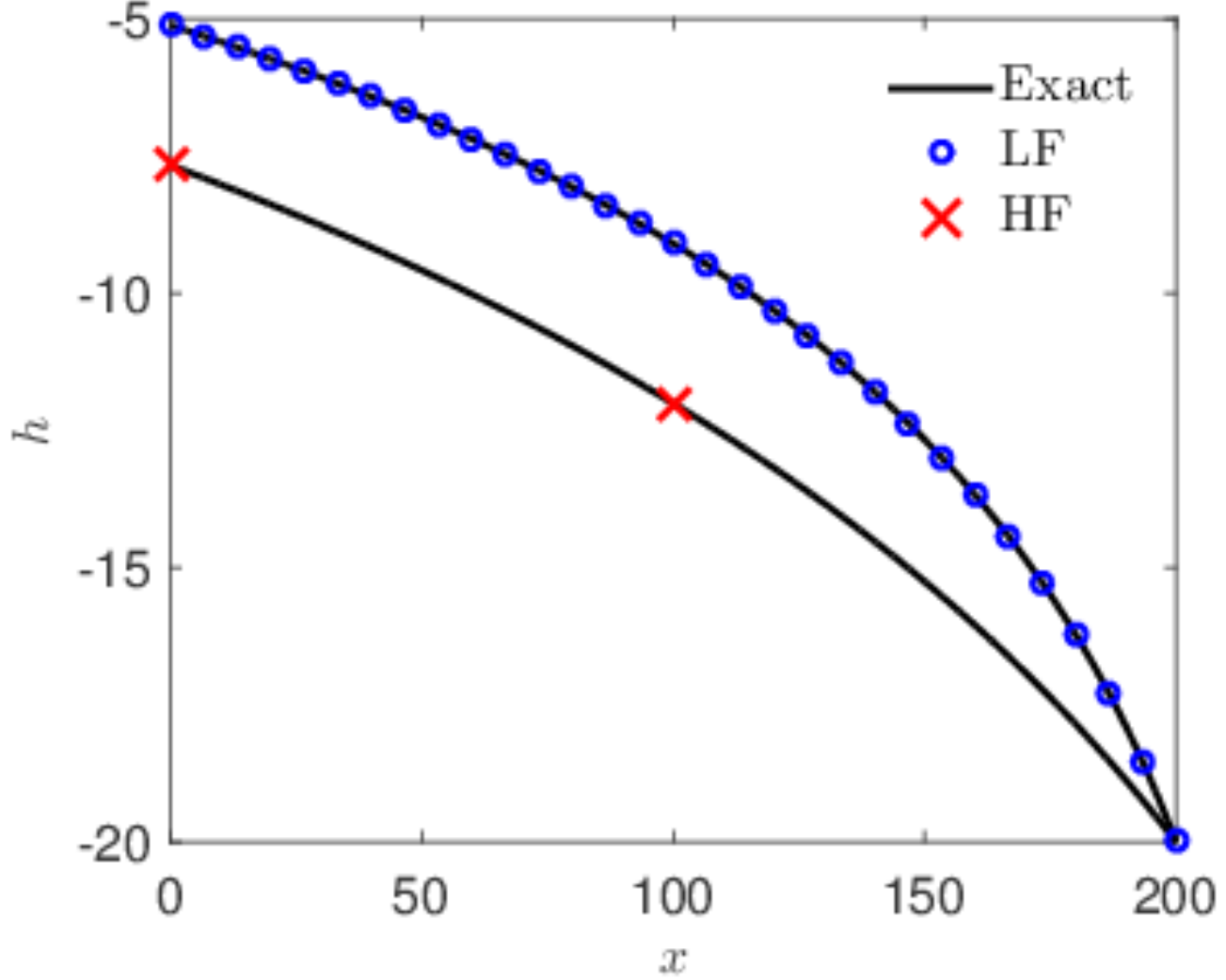}}
\subfigure[]{\label{difflhflowb}
\includegraphics[width = 0.45\textwidth]{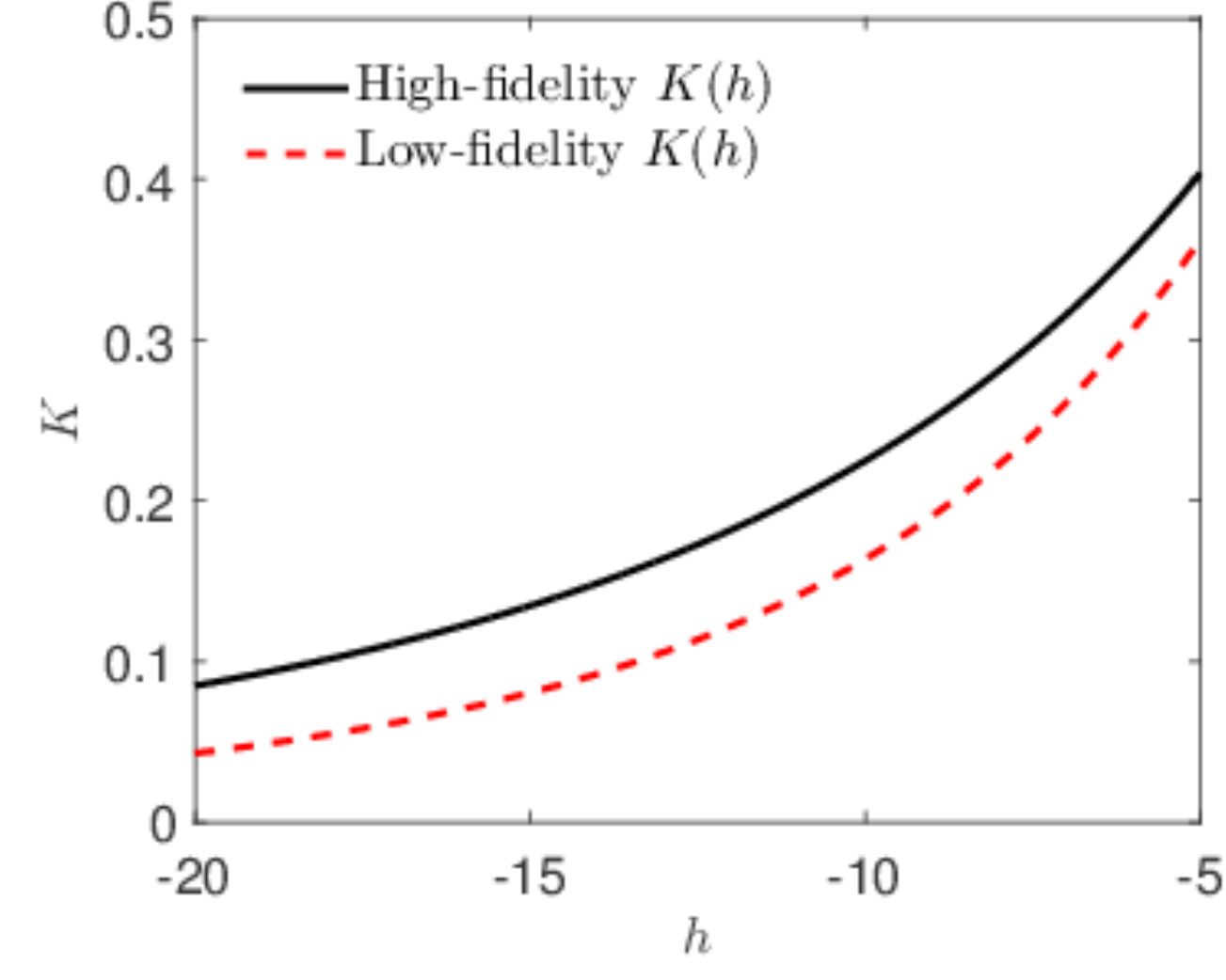}}
\subfigure[]{\label{diffflowa}
\includegraphics[width = 0.45\textwidth]{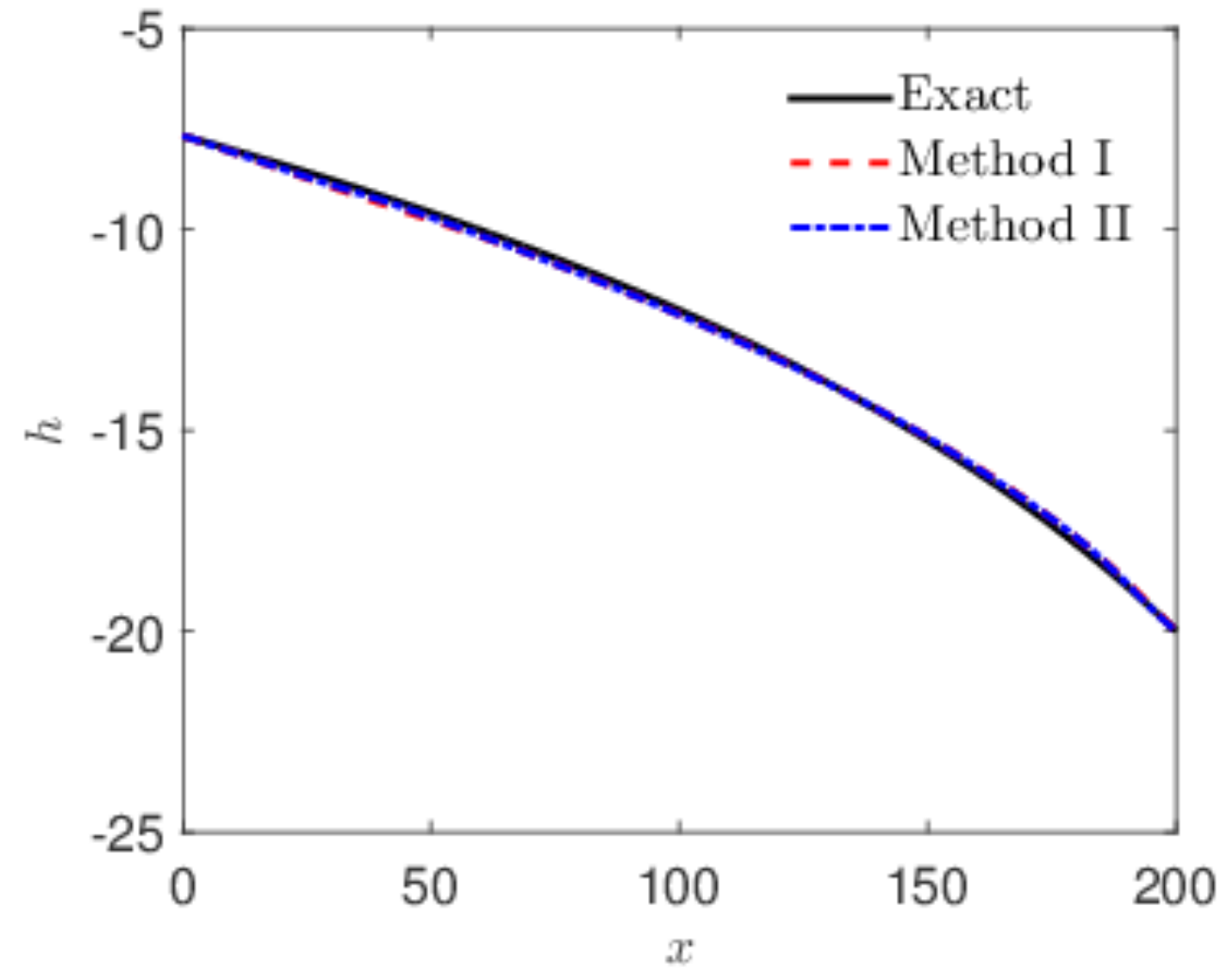}}
\subfigure[]{\label{diffflowb}
\includegraphics[width = 0.45\textwidth]{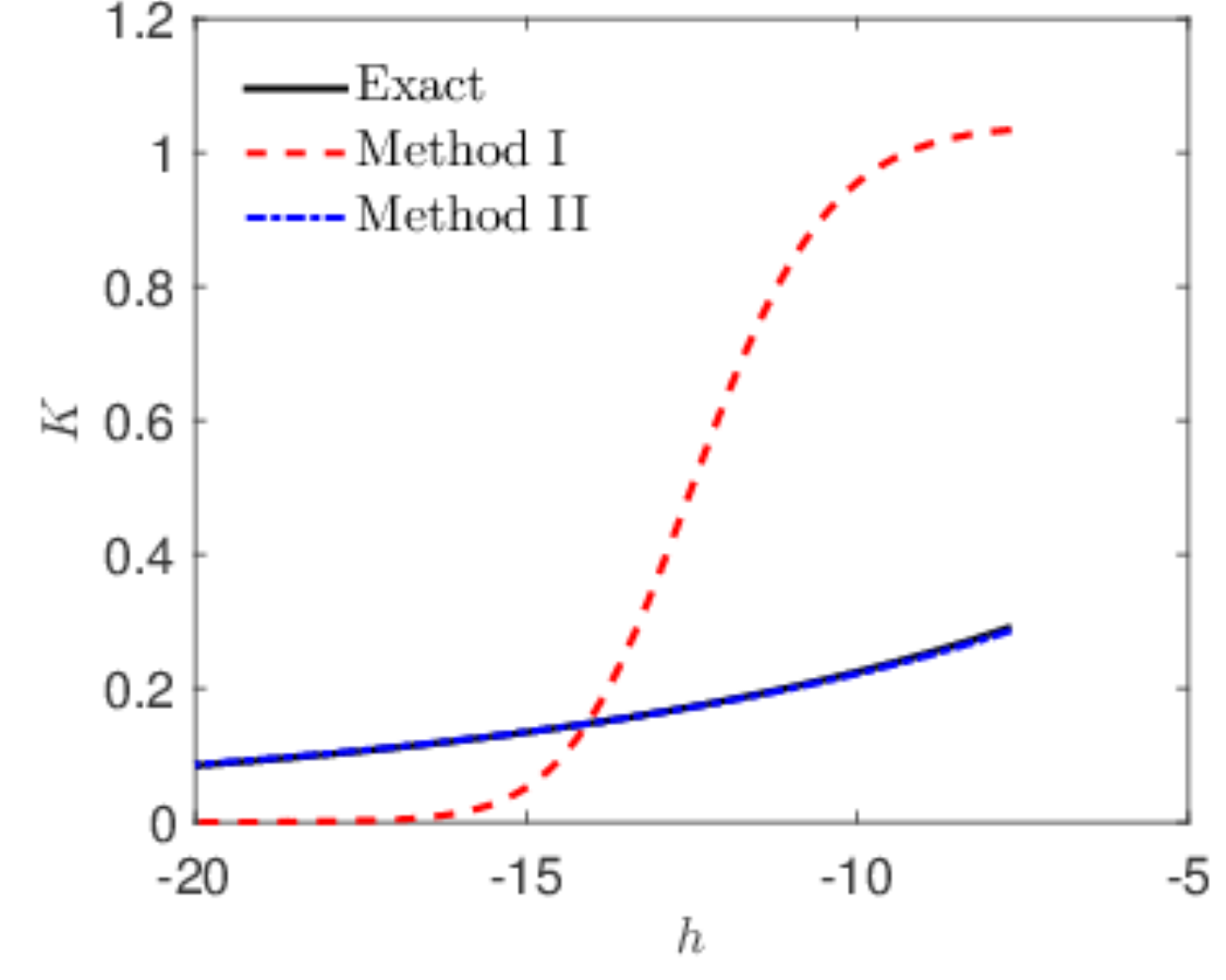}}
\caption{\label{diffflow} Predictions for unsaturated flow in porous media using the differential (Eq. \eqref{flows}) and integral formulations (Eq. \eqref{mass}) with constant flux at the inlet and constant pressure head at the outlet.  
(a) Training data for pressure head. 
(b) Low- and high-fidelity hydraulic conductivity. 
(c) Predicted pressure head using MPINNs training with multi-fidelity data. Method I: Differential formulation, Method II: Integral formulation. 
(d) Predicted hydraulic conductivity using MPINNs training with multi-fidelity data.  Method I: Differential formulation, Method II: Integral formulation. 
}
\end{figure}

\begin{figure}
\centering
\subfigure[]{\label{flowfluxa}
\includegraphics[width = 0.45\textwidth]{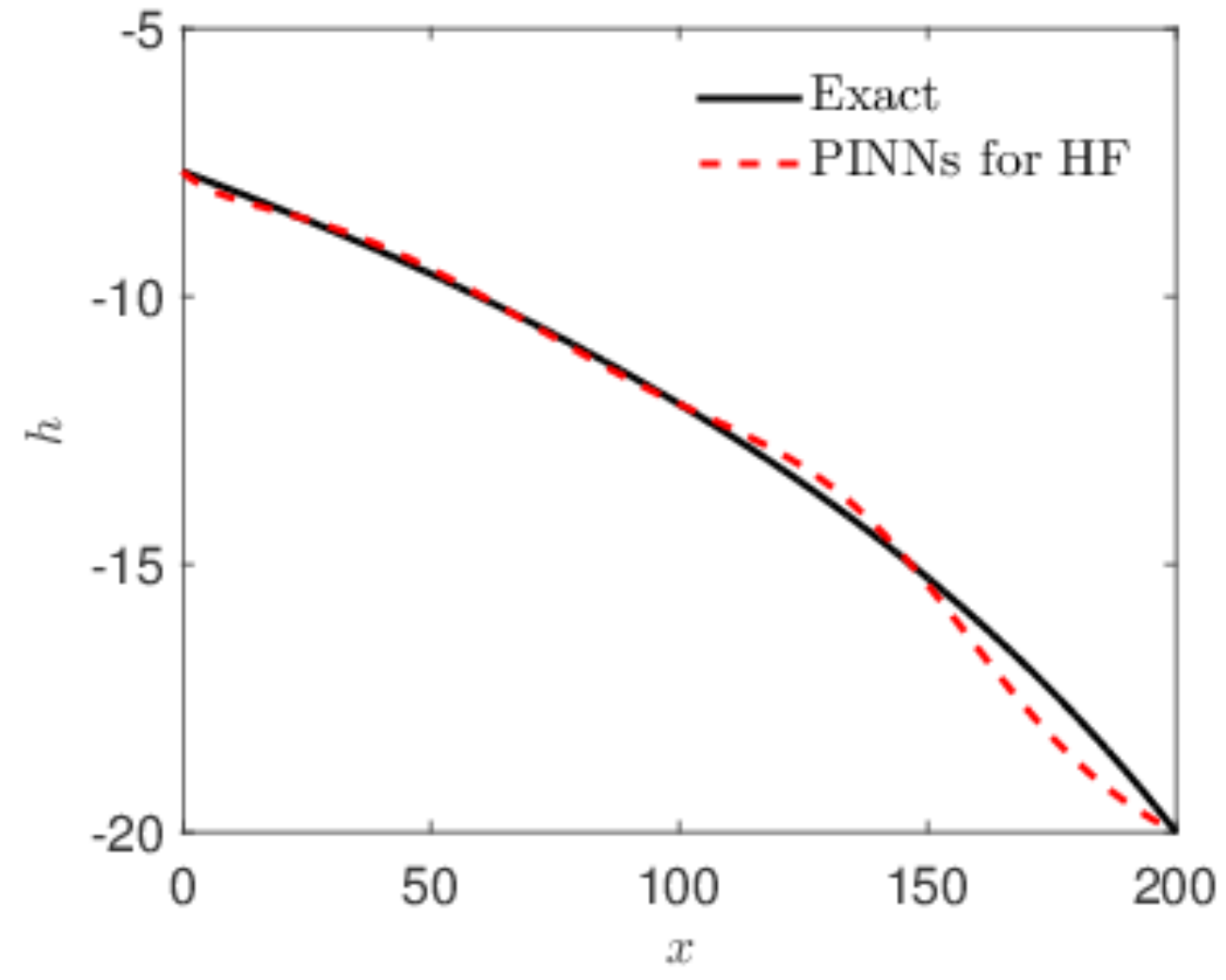}}
\subfigure[]{\label{flowfluxb}
\includegraphics[width = 0.45\textwidth]{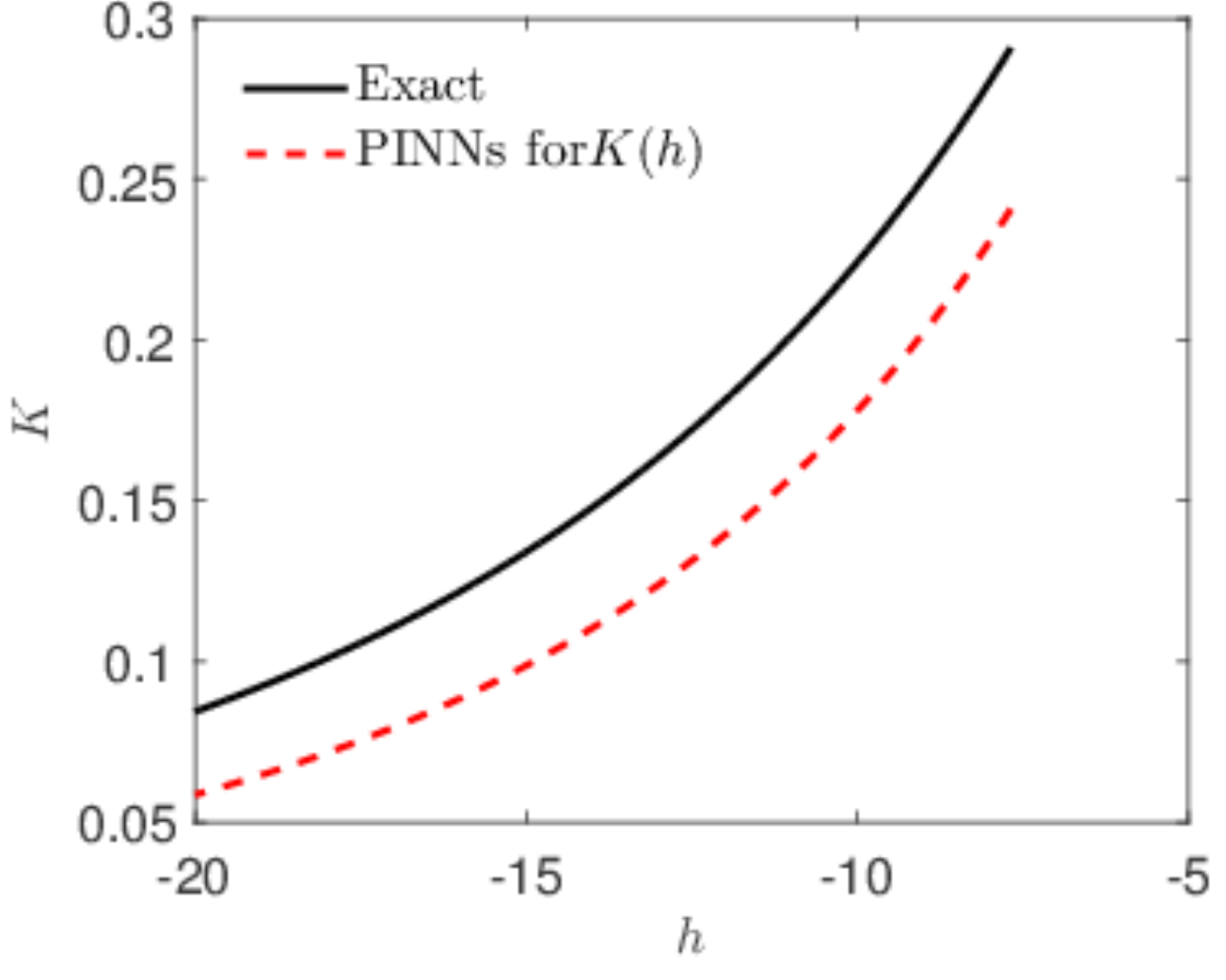}}
\subfigure[]{\label{flowfluxc}
\includegraphics[width = 0.45\textwidth]{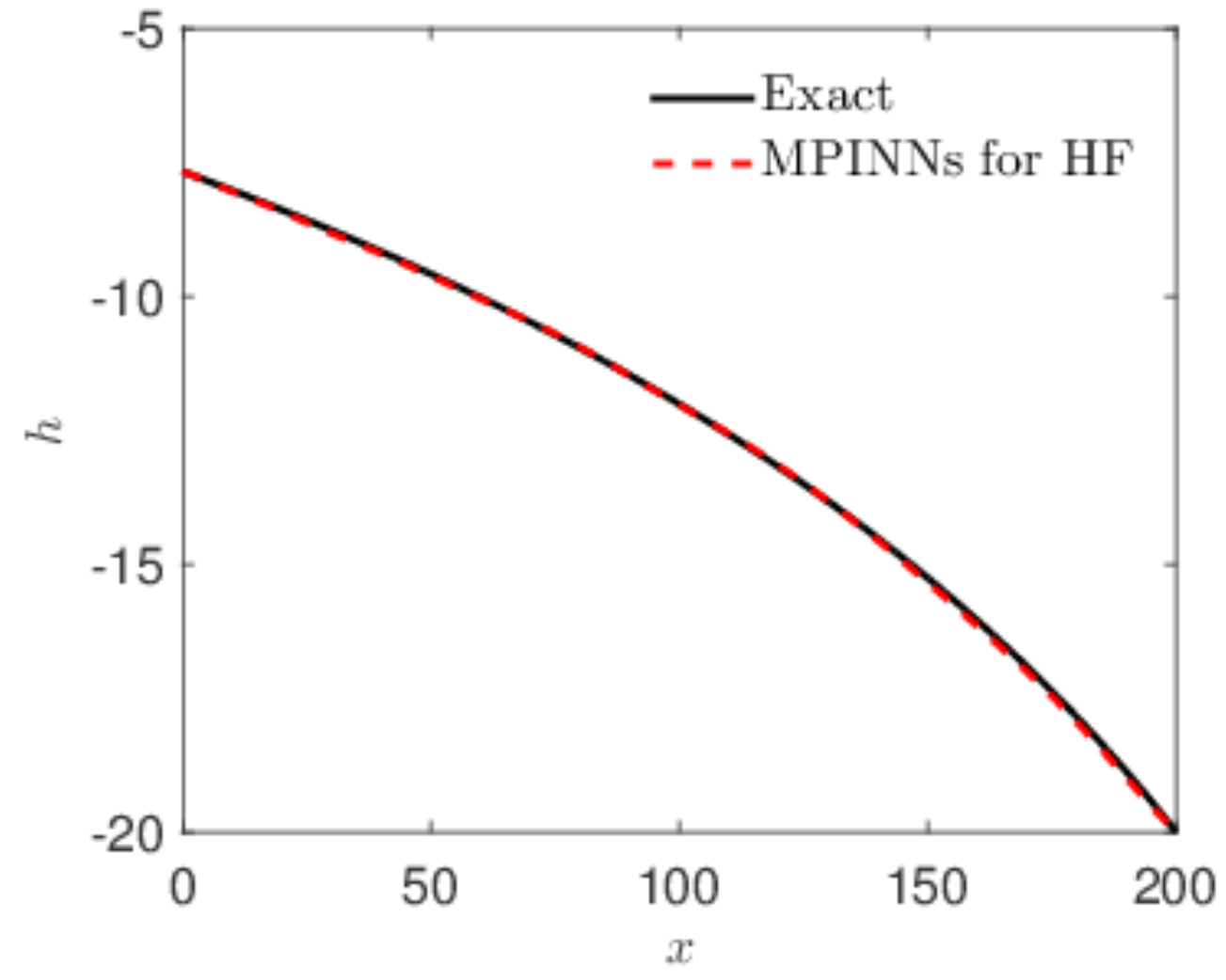}}
\subfigure[]{\label{flowfluxd}
\includegraphics[width = 0.45\textwidth]{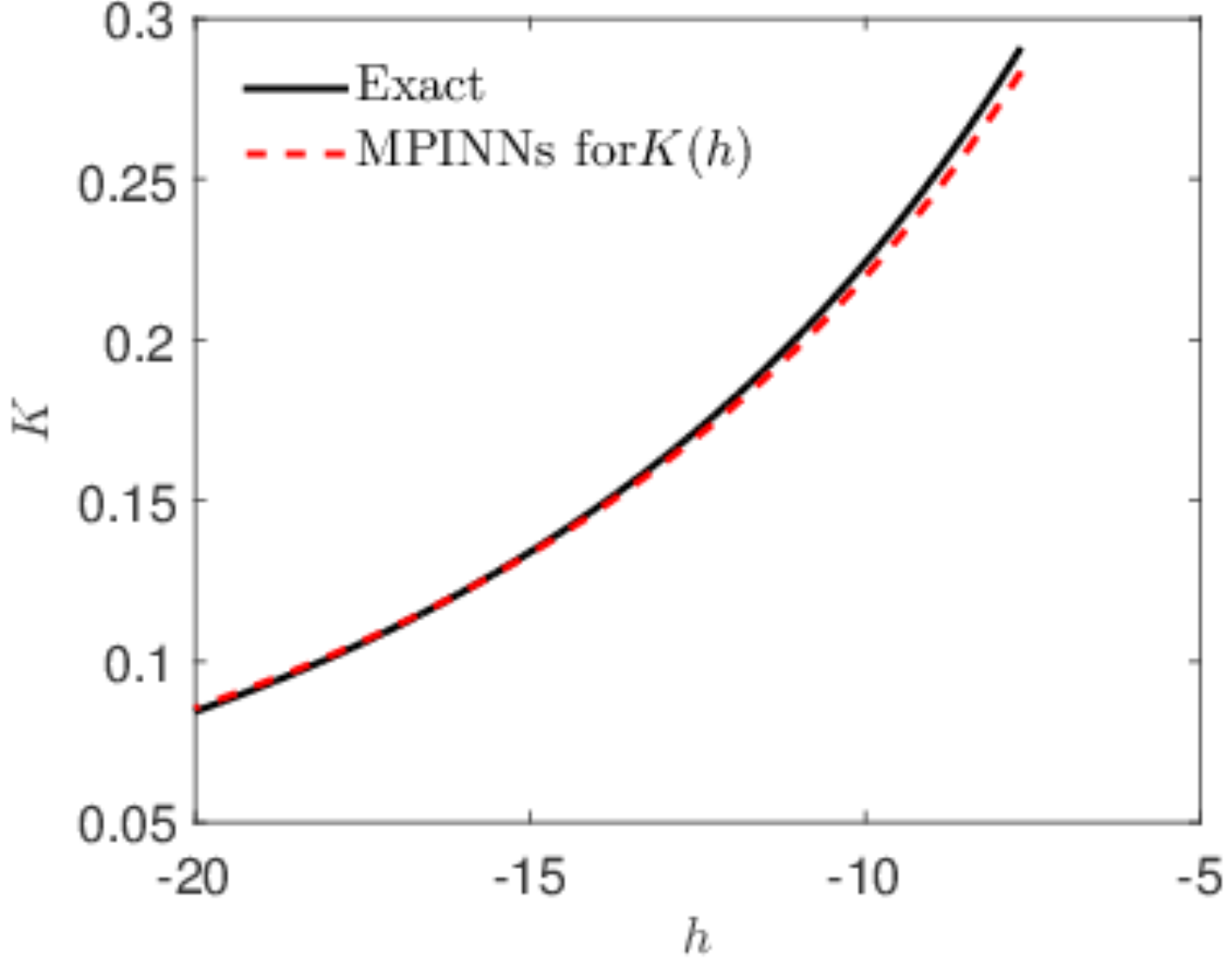}}
\caption{\label{flowflux} Predictions for unsaturated flow in porous media using the integral formulation (Eq. \eqref{mass}) with constant flux at the inlet and constant pressure head at the outlet.  
(a) Predicted pressure head using PINNs training with high-fidelity data only. 
(b) Predicted hydraulic conductivity using PINNs training with high-fidelity data only. 
(c) Predicted pressure head using MPINNs with multi-fidelity data. 
(d) Predicted hydraulic conductivity using using MPINNs with multi-fidelity data.
}
\end{figure}

The loam  is selected as a representative case here, for which the empirical ranges of $\alpha_0$ and $m$ are: $\alpha_0 (cm^{-1}) \in [0.015, 0.057]$ and $m \in [0.31, 0.40]$ \cite{carsel1988developing}. In addition, $K_s = 1.04 cm/hr$. To obtain the training data for neural networks, two types of numerical simulations are conducted to generate the low- and high-fidelity data using the {\sl bvp4c} in {\sl Matlab} (uniform lattice with $\delta_x = 1/15 cm$).  For high-fidelity data,  the exact values for $\alpha_0$ and $m$ are assumed to be 0.036 $cm^{-1}$ and 0.36. The high-fidelity simulations are then conducted using the exact values of $\alpha_0$ and $m$. Different initial guesses for $\alpha_0$ and $m$ are employed in the low-fidelity simulations. Specifically, ten uniformly distributed pairs i.e., $(\alpha_0, m)$ in the range $(0.015, 0.31) - (0.057, 0.40)$  are adopted in the low-fidelity simulations. For all cases, 31 uniformly distributed sampling data at the low-fidelity level are served as the training data, and only two sampling points including the boundary conditions mentioned above are employed as the training data for high-fidelity. In addition,  a smaller learning rate i.e., $10^{-4}$ is employed for all test cases in this section.

\begin{figure}
\centering
\subfigure[]{\label{lhflowa}
\includegraphics[width = 0.45\textwidth]{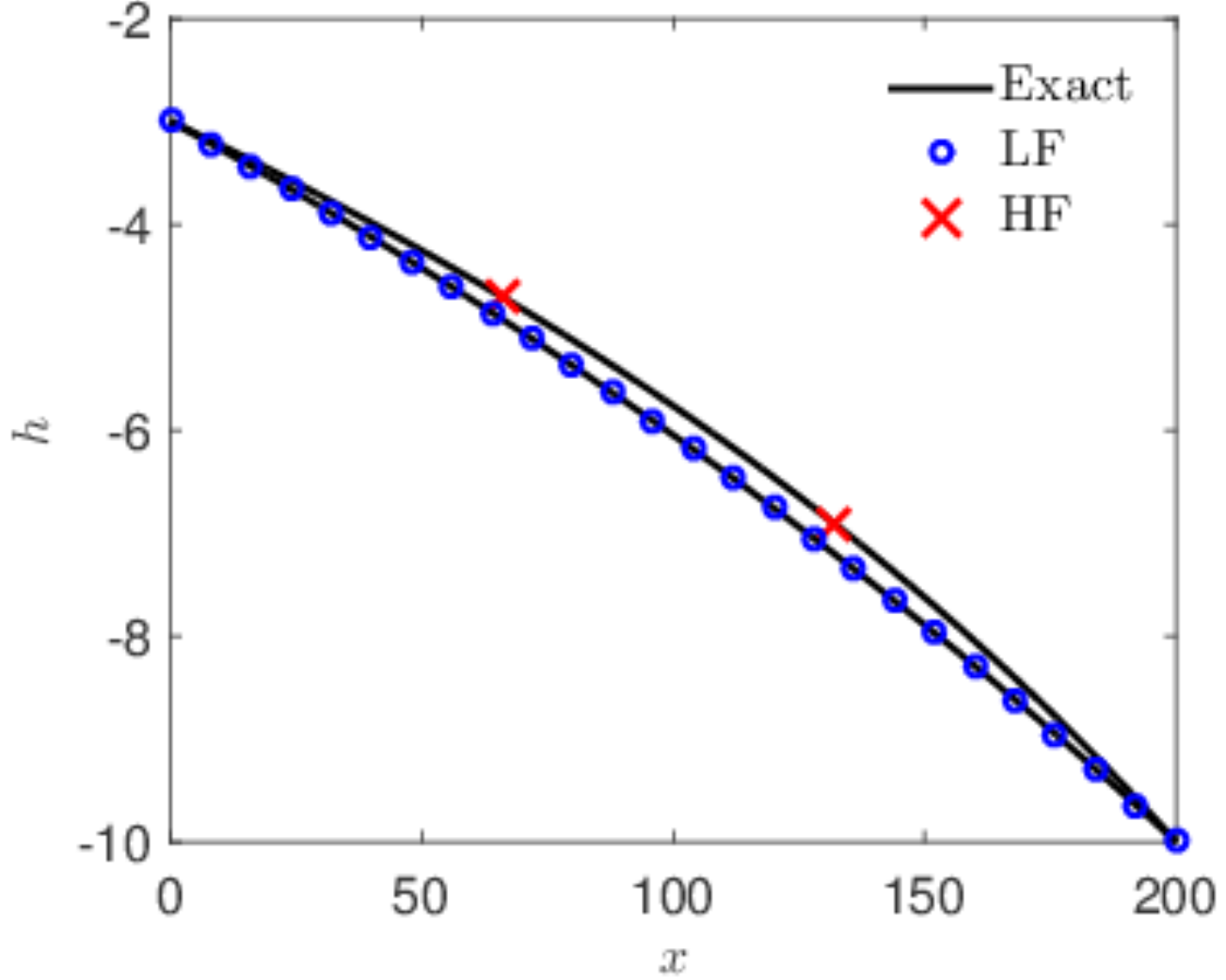}}
\subfigure[]{\label{lhflowb}
\includegraphics[width = 0.45\textwidth]{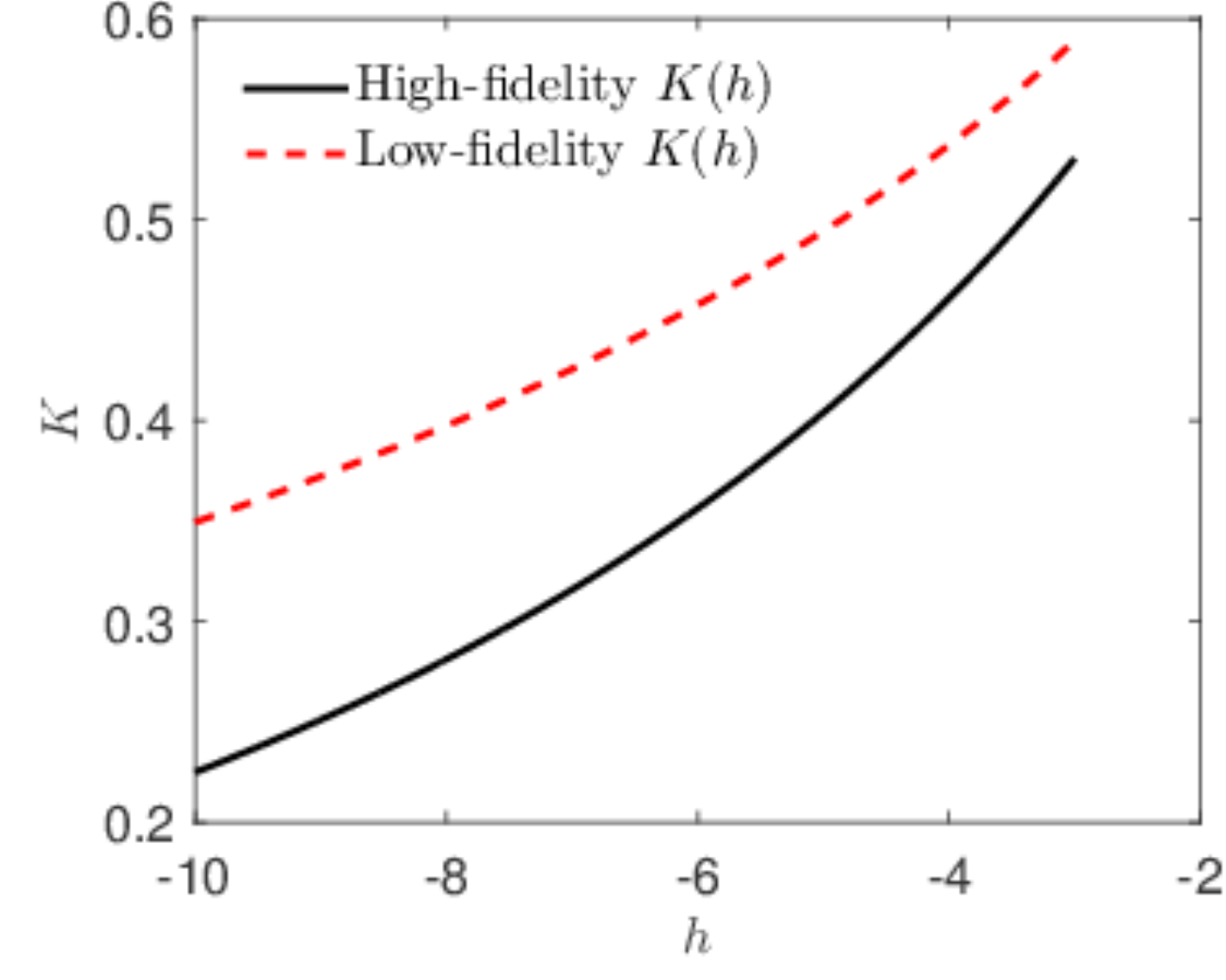}}
\subfigure[]{\label{flowprea}
\includegraphics[width = 0.45\textwidth]{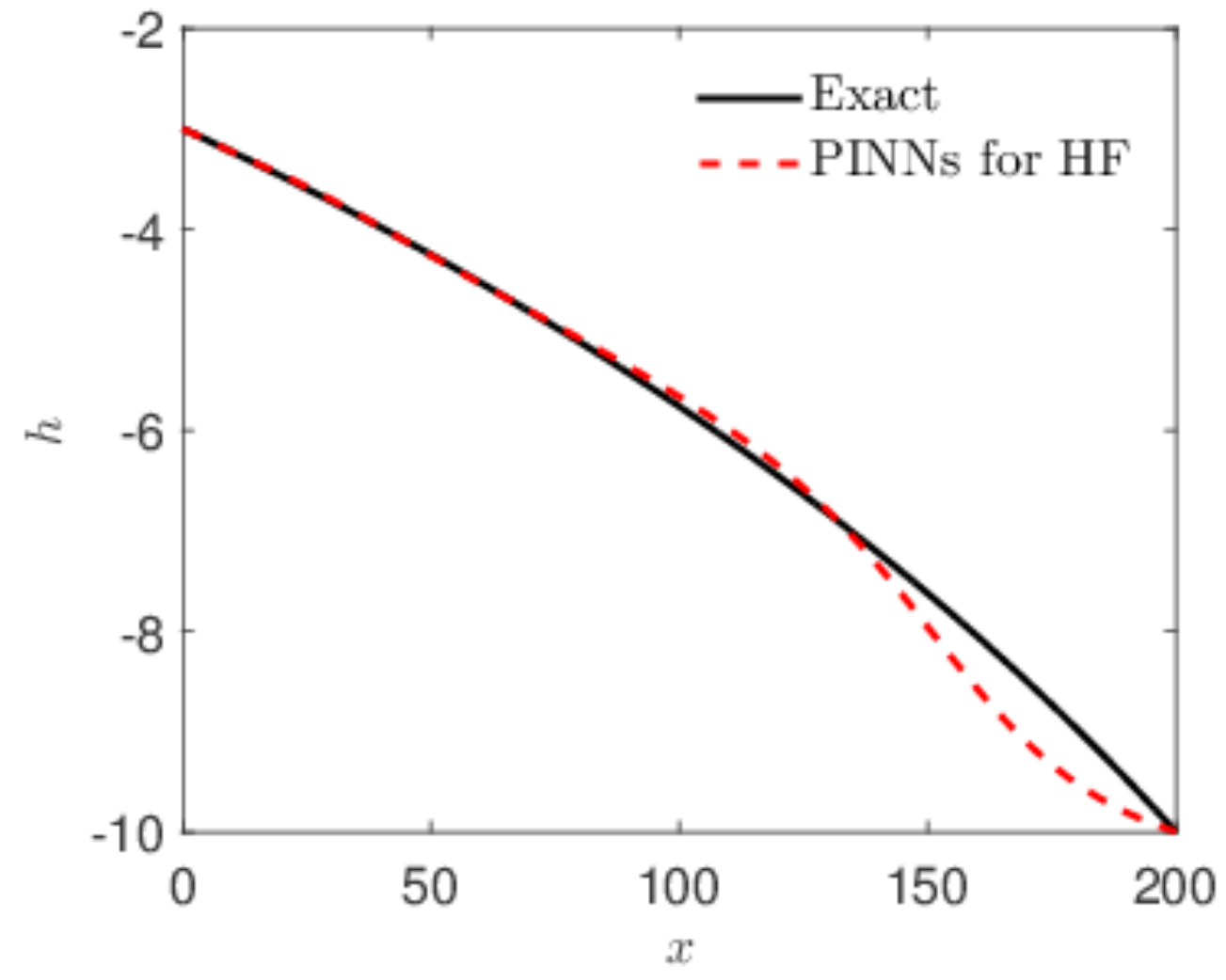}}
\subfigure[]{\label{flowpreb}
\includegraphics[width = 0.45\textwidth]{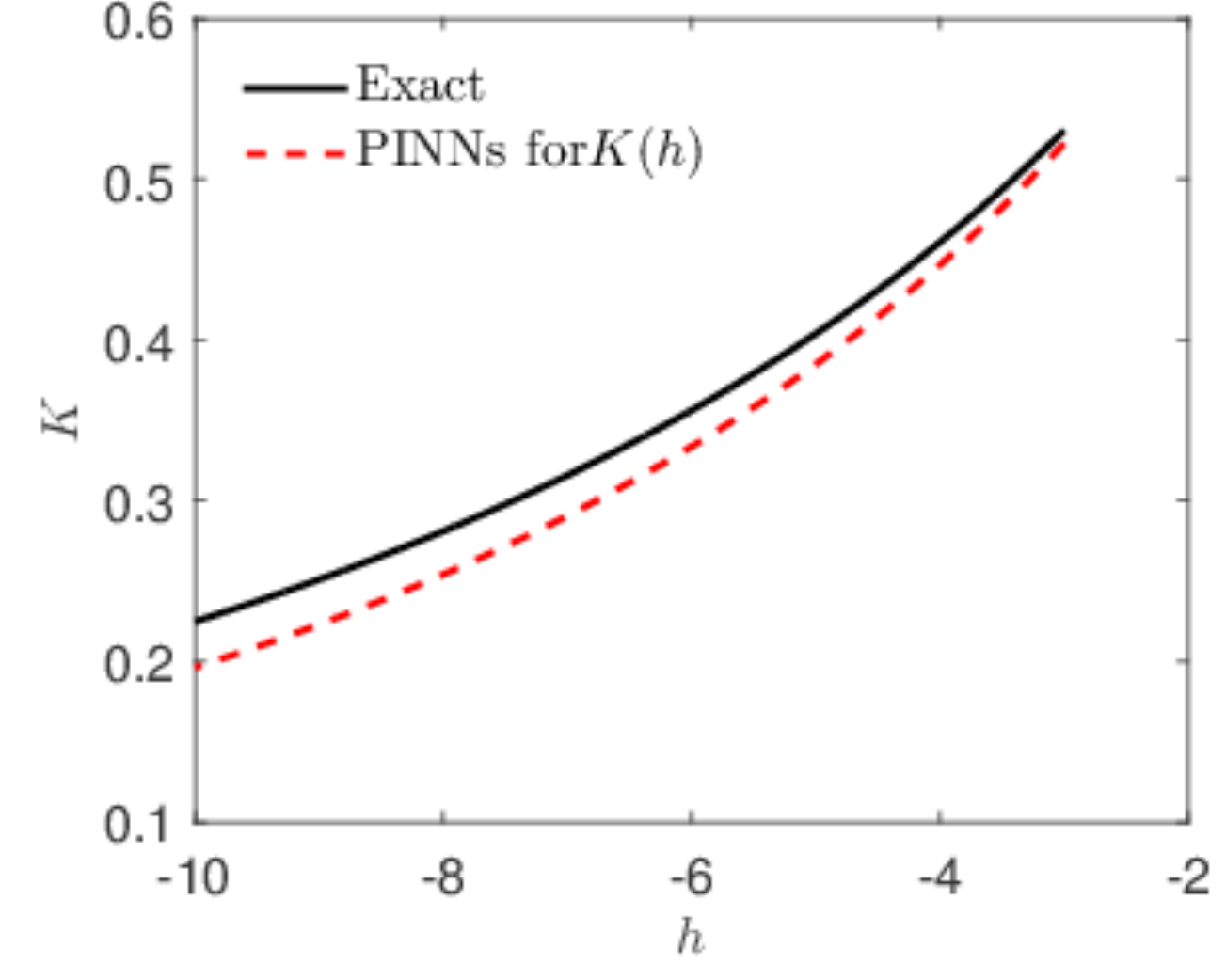}}
\subfigure[]{\label{flowprec}
\includegraphics[width = 0.45\textwidth]{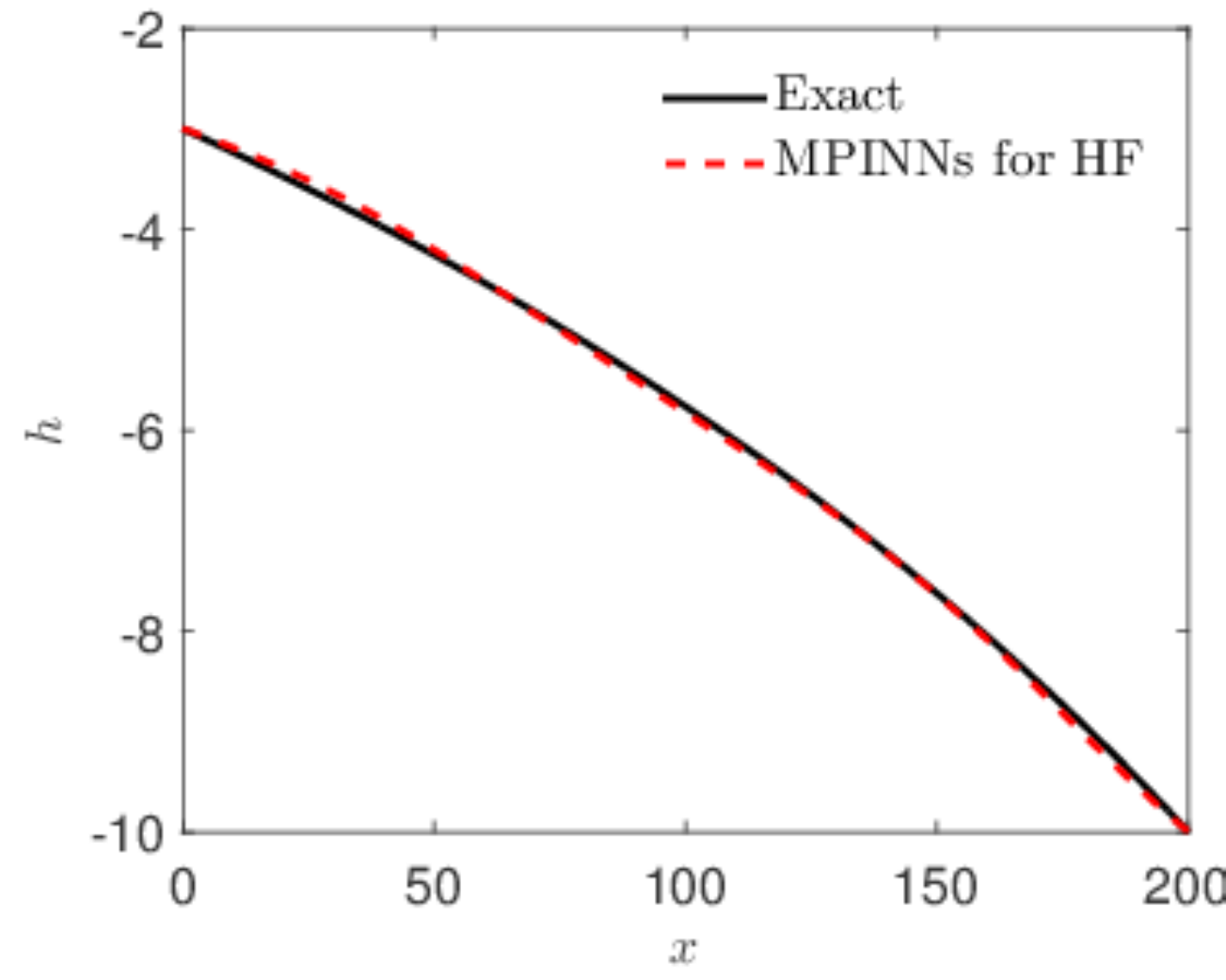}}
\subfigure[]{\label{flowpred}
\includegraphics[width = 0.45\textwidth]{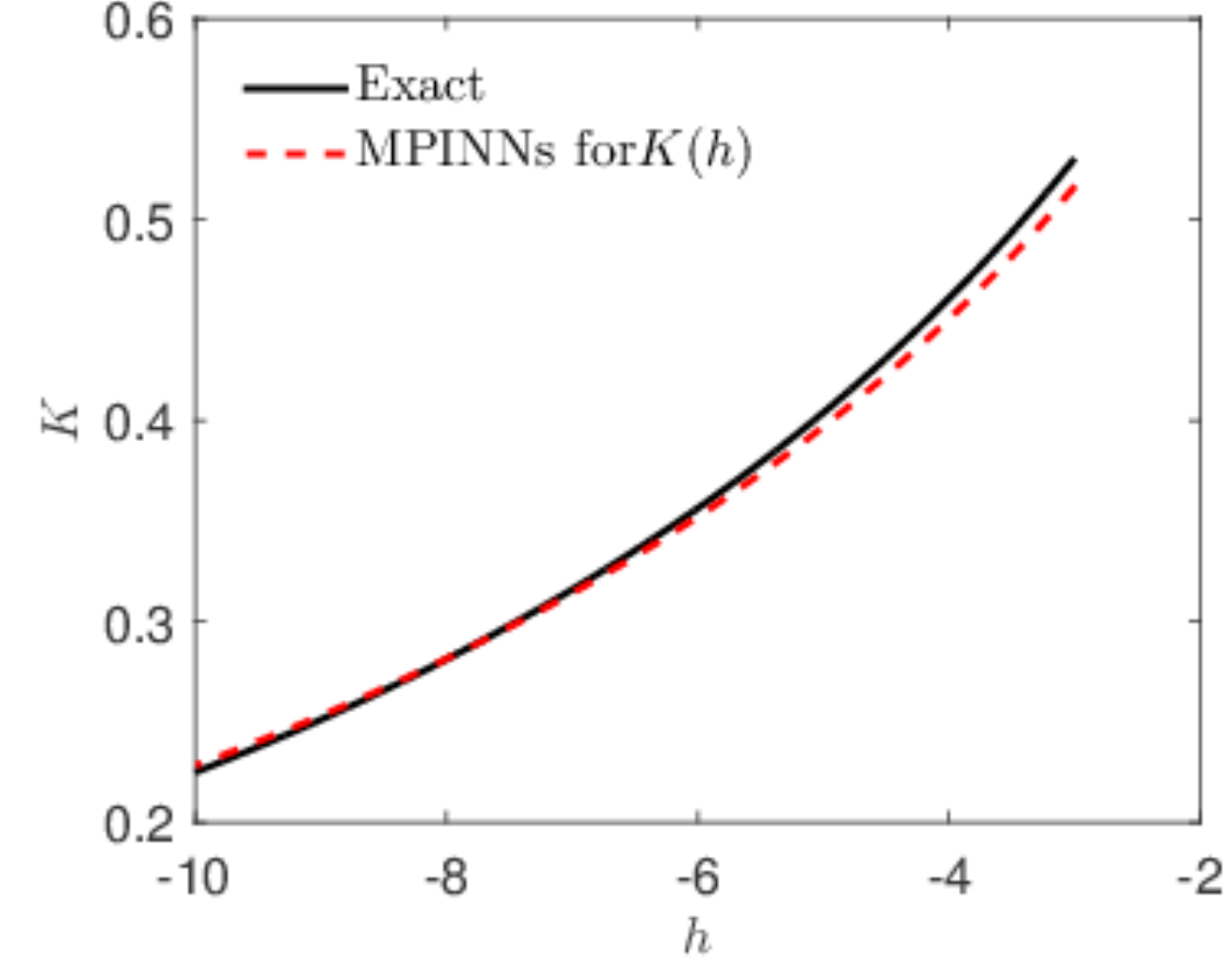}}
\caption{\label{flowpre} Predictions for unsaturated flow in porous media using the integral formulation (Eq. \eqref{mass}) with constant pressure head at the inlet and outlet.
(a) Training data for pressure head. Low-fidelity data is computed with $\alpha_0 = 0.015$ and $m = 0.31$.  
(b) Low- and high-fidelity hydraulic conductivity. Low-fidelity hydraulic conductivity is computed with $\alpha_0 = 0.015$ and $m = 0.31$.  
(c) Predicted pressure head using PINNs training with high-fidelity data only. 
(d) Predicted hydraulic conductivity using PINNs training with high-fidelity data only. 
(e) Predicted pressure head using MPINNs with multi-fidelity data. 
(f) Predicted hydraulic conductivity using using MPINNs with multi-fidelity data.
}
\end{figure}

We first consider the flow with constant flux inlet. The flux at the inlet and the pressure at the outlet are set as $q_0 = 0.01cm/y$ and $h_1 = -20 cm$, respectively. Equation \eqref{flows} is added into the last neural network in MPINNs. We employ the numerical results for $\alpha_0 = 0.055$ and $m = 0.4$ as the low-fidelity data. As shown in Fig. \ref{diffflowb}, the prediction for hydraulic conductivity is different from the exact solution. According to Darcy's law, we can rewrite Eq. \eqref{flows} as
\begin{align}
    q(x) = -K\partial_x h,~ \partial_x q(x) = 0.  
\end{align}
 Considering that $q = q_0$ at the inlet is a constant, we can then obtain the following equation
 \begin{align}\label{mass}
    q(x) = -K\partial_x h = q_0,
 \end{align}
which actually is the mass conservation at each cross section. We then employ Eq. \eqref{mass} instead of Eq. \eqref{flows} in the MPINNs, and the results improve greatly (Fig. \ref{diffflowb}).

We proceed to study this case in some more detail. We perform the single-fidelity modeling (SF) based on the high-fidelity data. We use two hidden layers with 20 neurons per layer in $\mathcal{NN}_{H_2}$, in which the hyperbolic tangent function is employed as the activation function. The learned pressure head and the hydraulic conductivity are shown in Figs. \ref{flowfluxa}-\ref{flowfluxb}. We observe that both the learned $h$ and $K(h)$ disagree with the exact results.  We then switch to  multi-fidelity modeling. Two hidden layers and 10 neurons per layer are used in $\mathcal{NN}_L$, and two hidden layers with 10 neurons per layer are utilized for $\mathcal{NN}_{H_2}$.  The predicted pressure head as well as the hydraulic conductivity (average value from ten runs with different initial guesses) agree quite well with the exact values (Figs. \ref{flowfluxc}-\ref{flowfluxd}).
For Case II, we set the pressure head at the inlet and outlet as $h_0 = -3 cm$ and $h_1 = -10 cm$. We also assume that the flux at the inlet is known, thus Eq. \eqref{mass} can  also be employed instead of Eq. \eqref{flows} in the MPINNs. The training data are illustrated in Fig. \ref{lhflowa}. The size of the NNs here is kept the same as that used in Case I. We observe that results for the present case (Figs. \ref{flowprea}-\ref{flowpred}) are quite similar with those in Case I.

Finally, the mean values of $\alpha_0$ as well as the $m$ for different initial guesses are shown in Table \ref{l2}, which indicates that the MPINNs can significantly improve the prediction accuracy as compared to the estimations based on the high-fidelity only (SF in Table \ref{l2}).

\begin{table}[htbp]
\centering
 \caption{\label{l2}PINN and MPINN predictions for hydraulic conductivity.}
 \begin{tabular}{ccccc}
  \hline \hline
  ~ & $\alpha_0 (cm^{-1})$ & $\sigma(\alpha_0)$ & $m$ & $\sigma(m)$ \\ \hline
  SF (Case I) & $0.0438 $ &- & $0.359$ &-\\
 MF (Case I) & $0.0344$ & $0.0027$ & 0.347& 0.0178\\
 SF (Case II) & $0.0440 $ &- & $0.377$ &-\\
 MF (Case II) & $0.0337$ & $7.91 \times 10^{-4}$ & 0.349& 0.0037\\
 Exact  & $0.036 $ &- & $0.36$ &-\\
  \hline \hline
 \end{tabular}
\end{table}

\subsubsection{Estimation of reaction models for reactive transport}
We further consider a single irreversible chemical reaction in a 1D soil column with a  length of 5$m$, which is expressed as 
\begin{align}
a_r A \rightarrow B,
\end{align}
where $A$, and $B$ are different solute. The above reactive transport can be described by the following advection-dispersion-reaction equation as
\begin{align}
\partial_t (\psi C_i) + q  \partial_x C_i =  \psi D\partial^2_x C_i - \psi v_i k_{f,r} C_A^{a_r}, (i = A, B),
\end{align}
where $C_i (mol/L)$ is the concentration of any solute, $q$ is the Darcy velocity, $\psi$ is the porosity,  $D$ is the dispersion coefficient, $k_{f,r}$ denotes the chemical reaction rate, $a_r$ is the order of the chemical reaction, both of which are difficult to measure directly, and $v_i$ is the stoichiometric coefficient with $v_A = a_r$, and $v_B = -1$.  Here, we assume that the following parameters are known:  $\psi = 0.4$, $q = 0.5m/y$,  and $D = 10^{-8} m/s^2$. The initial and boundary conditions imposed on the solute are expressed as
\begin{align}
C_A(x, 0) = C_B(x, 0)  = 0,\\
C_A(0, t)  = 1, ~C_B(0, t) = 0,\\
\partial_x C_i(x, t)|_{x =l_x} = 0.
\end{align} 
The objective here is to learn the effective chemical reaction rate as well as the reaction order based on partial observations of the concentration field $C_A(x, t)$.

\begin{figure}
\centering
\subfigure[]{\label{ctraina}
\includegraphics[width = 0.45\textwidth]{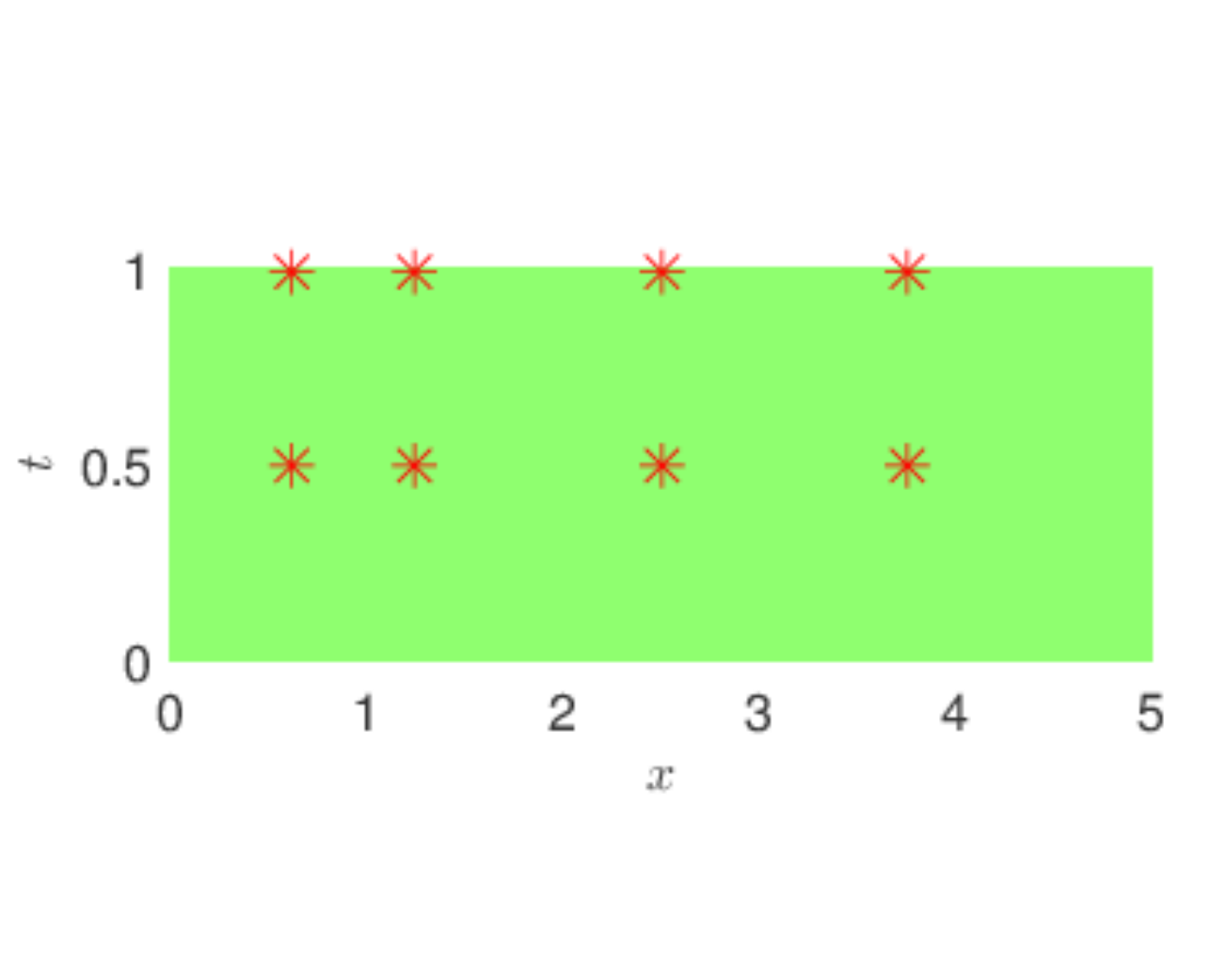}}
\subfigure[]{\label{ctrainb}
\includegraphics[width = 0.45\textwidth]{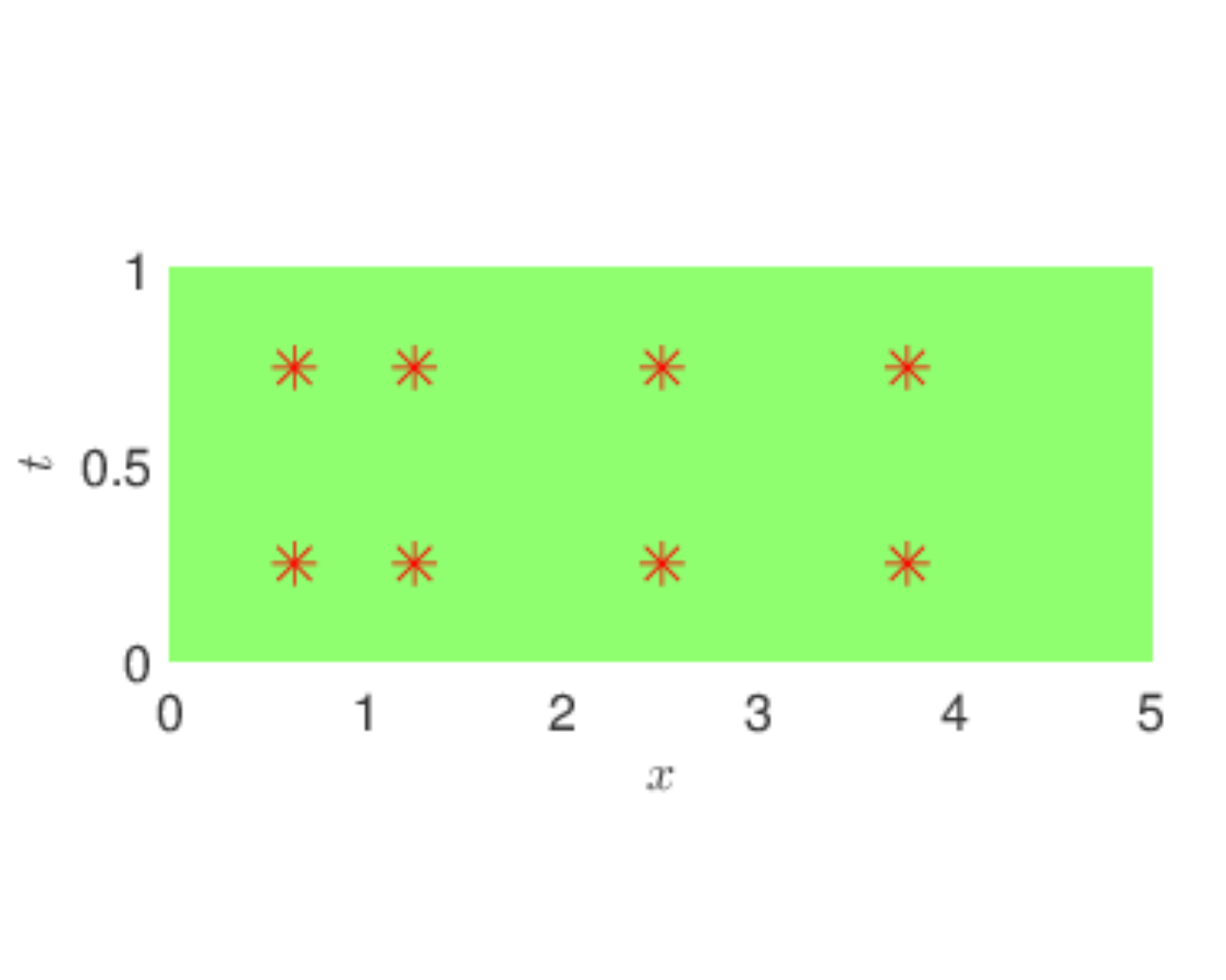}}
\caption{\label{ctrain} Schematic of the space-time domain and the locations of the  high-fidelity data for  modeling reactive transport. (a) Case I: Data are collected at $t = 0.5$ and 1 years. (b) Case II: Data are collected at $t = 0.25$ and 0.75 years.}
\end{figure}

We perform lattice Boltzmann simulations \cite{meng2016localized,shi2009lattice} to obtain the training data since we have no experimental data. Consider that $v_A$ is a constant, we define an effective reaction rate as $k_f = v_A k_{f,r}$ for simplicity.  The exact effective reaction rate and reaction order are assumed to be $k_f = 1.577 / y$ and $a_r = 2$, respectively. Numerical simulations  with the exact $k_f$ and $a_r$ are then conducted to obtain the high-fidelity data. In simulations, a uniform lattice is employed, i.e., $l_x = 400 \delta_x$, where $\delta_x = 0.0125m$ is the space step, and $\delta_t = 6.67 \times 10^{-4} y$ is the time step size. We assume that the sensors for concentration are located at $x = \{0.625, 1.25, 2.5, 3.75\}m$. In addition, we assume that the data are collected from the sensors once half a year. In particular, we employ two different datasets (Fig. \ref{ctrain}), i.e., (1) $t = 0.5$ and 1 years (Case I), and (2) $t = 0.25$ and 0.75 years.
Schematics of the training data points for the two cases we consider are shown in Fig. \ref{ctraina}
and Fig. \ref{ctrainb}.

Next, we describe how we obtain the low-fidelity data. In realistic applications, the pure chemical reaction rate (without porous media)  between different solute e.g., $A$ and $B$ are known, which can be served as the initial guess for $k_f$. Here we assume that the initial guess for the chemical reaction rate and reaction order vary from $0.75 k_f/a_r$ to $1.25 k_f/a_r$. To study the effect of the initial guess ($(k_{f,0}, a_{r0})$) on the predictions, we conduct the low-fidelity numerical simulations based on ten uniformly distributed pairs in $[0.75 k_f, 0.75a_r] - [1.25k_f, 1.25a_r]$ using the same grid size and time step as the high-fidelity simulations. Here $k_{f,0}$ and $a_{r0}$ represent the initial guesses for $k_f$ and $a_r$. The learning rate employed in this section is also $10^{-4}$.


\begin{figure}
\centering
\subfigure[]{\label{reactiona}
\includegraphics[width = 0.45\textwidth]{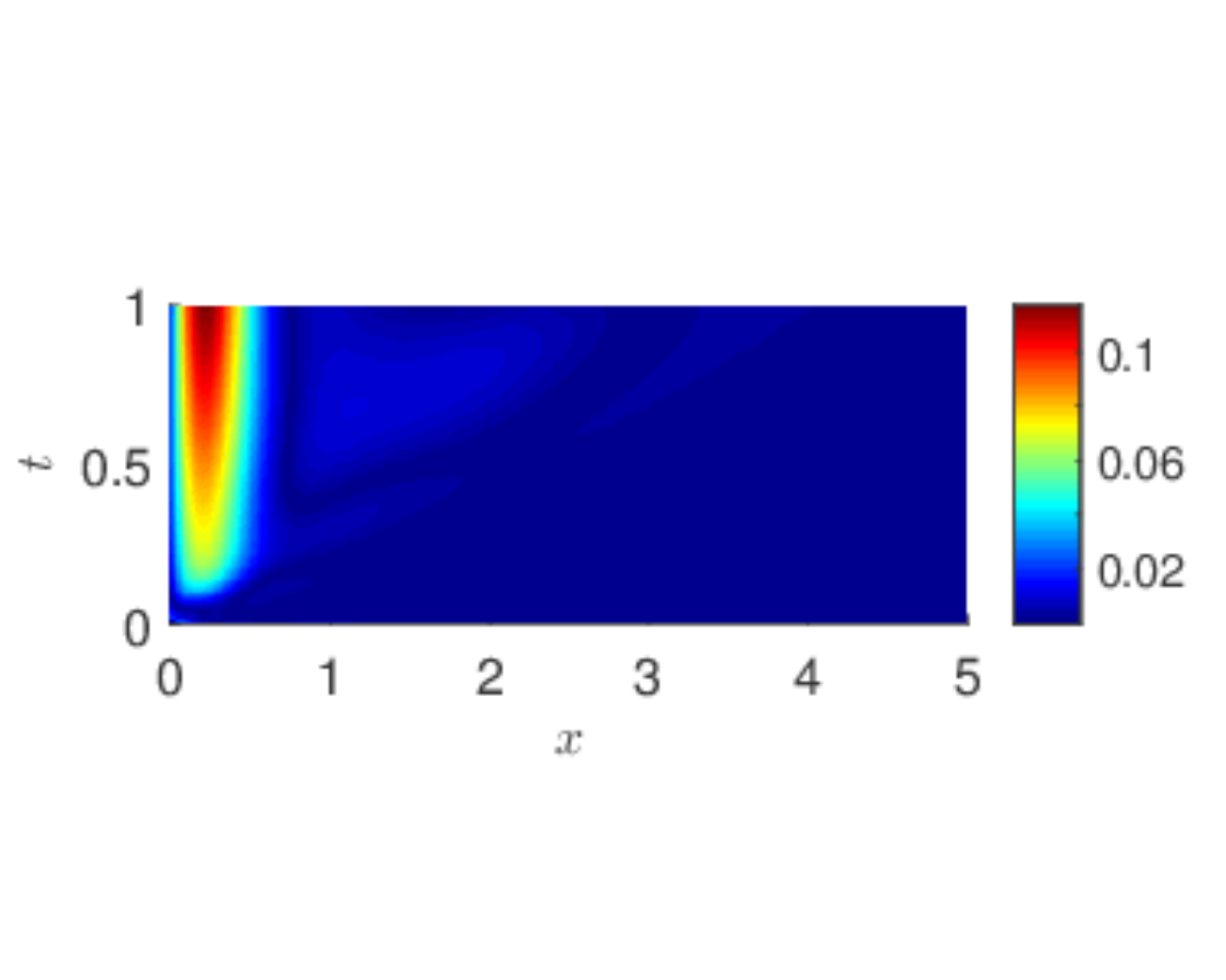}}
\subfigure[]{\label{reactionb}
\includegraphics[width = 0.45\textwidth]{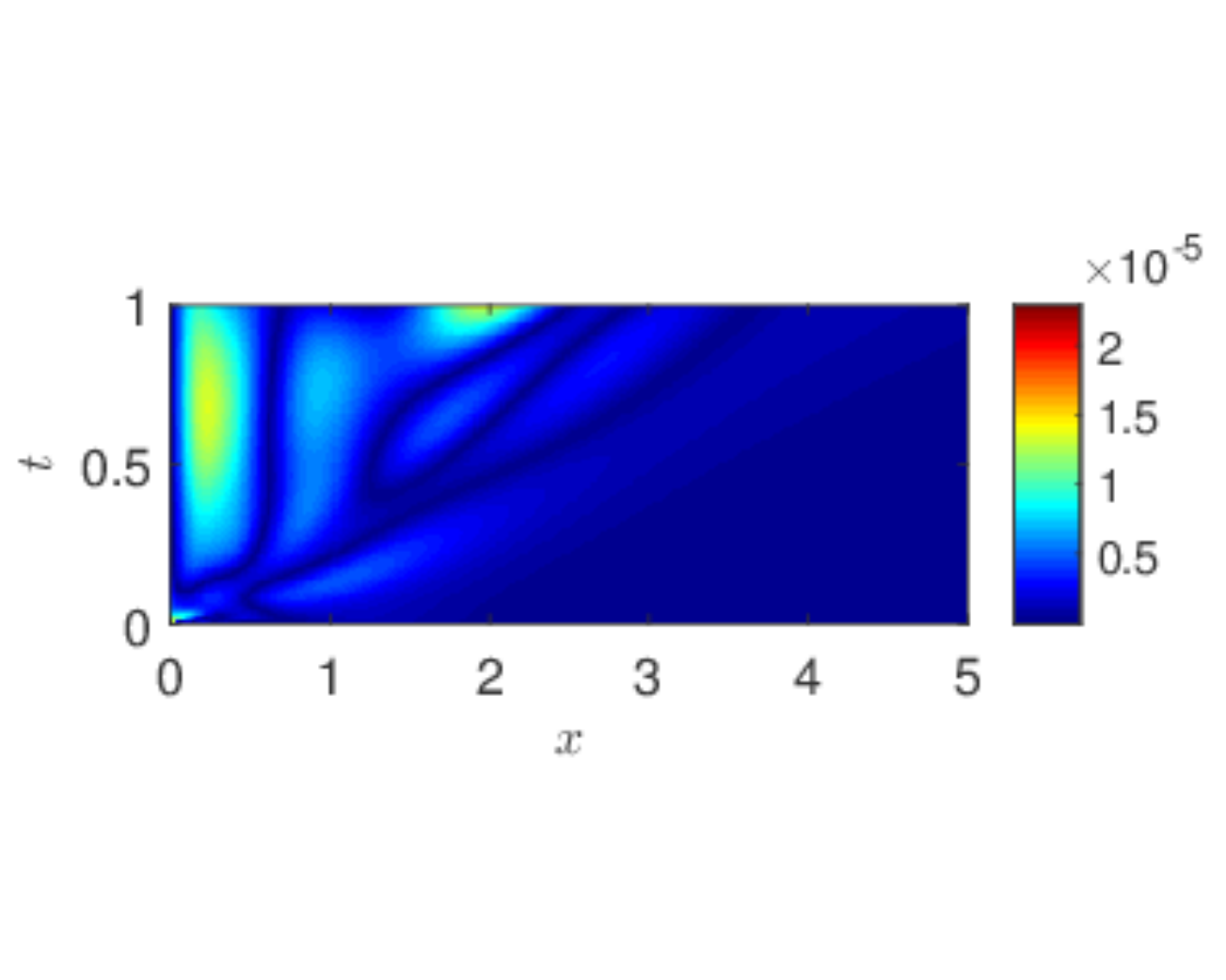}}
\subfigure[]{\label{reactionc}
\includegraphics[width = 0.45\textwidth]{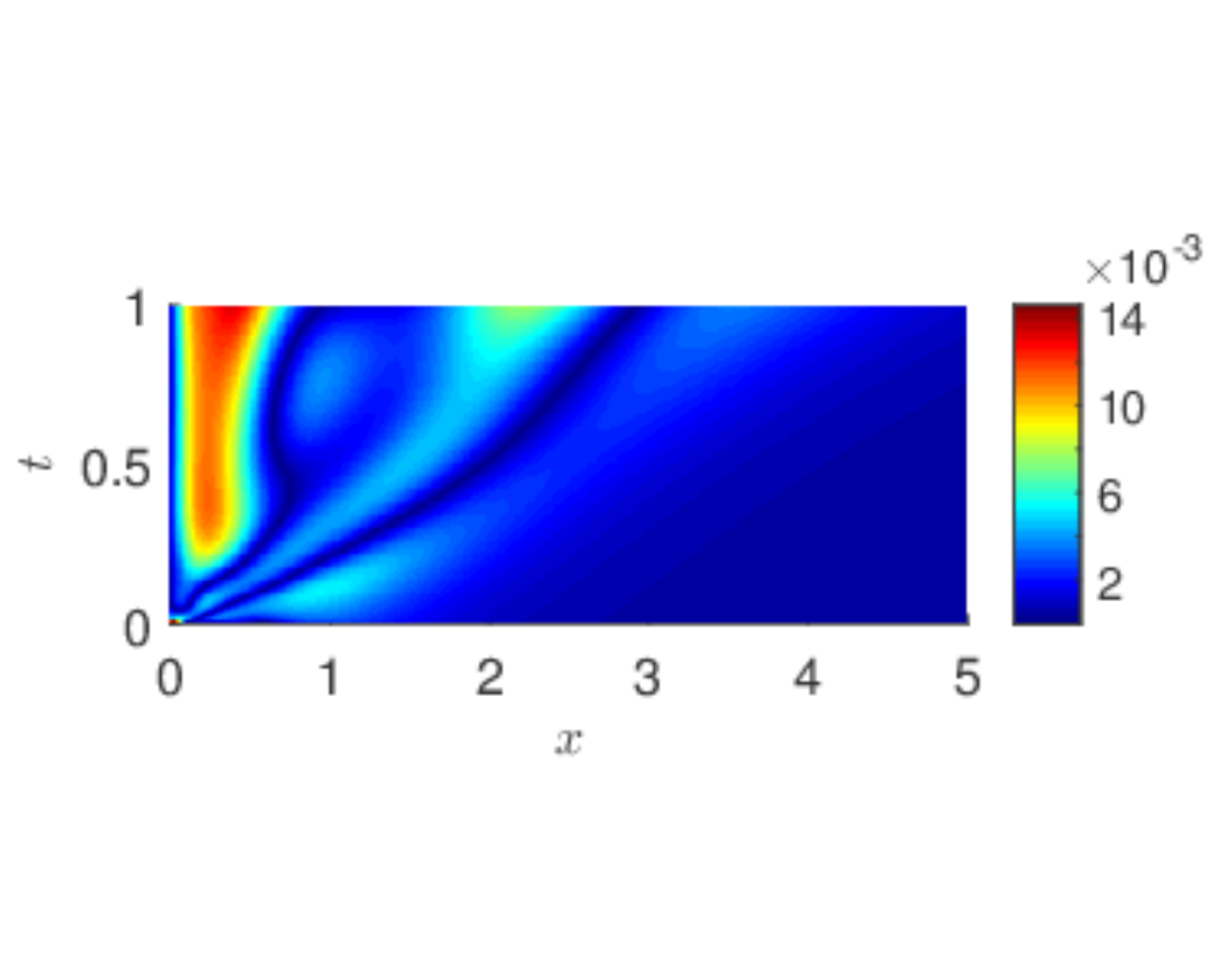}}
\subfigure[]{\label{reactiond}
\includegraphics[width = 0.45\textwidth]{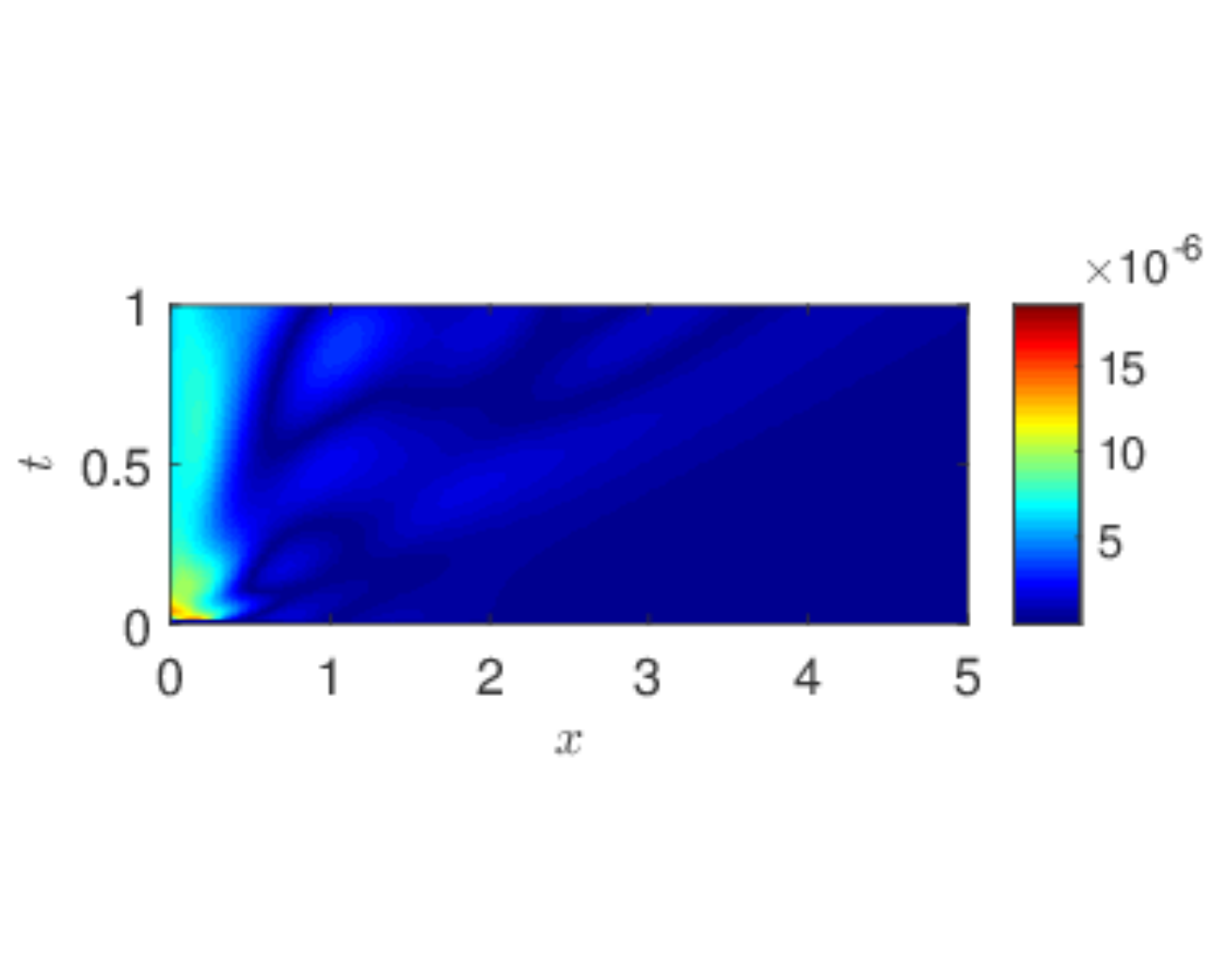}}
\caption{\label{reaction} Predicted concentration field.
(a) Case I: Relative errors (absolute value)  using a PINN trained on high-fidelity data only. $\mathcal{NN}_{H_2} \rightarrow 4 \times 20$. 
(b) Case I: Mean relative errors (absolute value) using a MPINN trained on multi-fidelity data. Initial guesses: ten uniformly distributed pairs in $[0.75 k_f, 0.75a] - [1.25k_f, 1.25a]$. The concentration fields plotted are the mean values for ten runs with different initial guesses. $\mathcal{NN}_{L} \rightarrow 2 \times 10$, $\mathcal{NN}_{H_2} \rightarrow 2 \times 10$.
(c) Case II: Relative errors (absolute value)  using a PINN trained on high-fidelity data only. $\mathcal{NN}_{H_2} \rightarrow 4 \times 20$.
(d) Case II: Mean relative errors (absolute value) using a MPINN trained on multi-fidelity data.  $\mathcal{NN}_{L} \rightarrow 2 \times 10$, $\mathcal{NN}_{H_2} \rightarrow 2 \times 10$.
}
\end{figure}

The results of predictions using PINNs (with the hyperbolic tangent activation function) trained on high-fidelity data are shown in Fig. 
\ref{reactiona} and Fig. \ref{reactionc} for the two cases we consider, and corresponding results 
using MPINNs are shown in Fig. \ref{reactionb} and Fig. \ref{reactiond}.
The estimated mean and standard deviation for $k_f$ and $a_r$ are displayed in Table \ref{reactiontb}, which are much better than the results from single-fidelity modelings. We  also note that the standard deviations are rather small, which demonstrates the robustness of the MPINNs. 

\begin{table}[htbp]
\centering
 \caption{\label{reactiontb}PINN and MPINN predictions for reactive transport.}
 \begin{tabular}{ccccc}
  \hline \hline
   & $k_{f} (/y)$ & $\sigma(k_{f})$ & $a_r$ & $\sigma(a_r)$ \\ \hline
  SF (Case I) & 0.441  & -  & 0.558 & - \\
  MF (Case I) & 1.414  & $7.45 \times 10^{-3}$  & 1.790 & $9.44\times 10^{-3}$ \\
  SF (Case II) & 1.224  & -  & 1.516 & - \\
  MF (Case II) & 1.557  & $2.14 \times 10^{-2}$  & 1.960 & $2.57\times 10^{-2}$ \\
  Exact & 1.577  & -  & 2 & - \\
  \hline \hline
 \end{tabular}
\end{table}

\section{Conclusion}
\label{summary}
In this work we presented a new composite deep neural network that learns from multi-fidelity data,
i.e. a small set of high-fidelity data and a larger set of inexpensive low-fidelity data. This scenario
is prevalent in many cases for modeling physical and biological systems and we expect that the
new DNN will provide solutions to many current bottlenecks where availability of large data sets of high-fidelity is simply not possible but either low-fidelity data from inexpensive sensors or other modalities or even simulated data can be readily obtained. Moreover, we extended the concept of
physics-informed neural networks (PINNs) that use a single-fidelity data to train to the multi-fidelity case and MPINNs. Specifically,  MPINNs are composed of four fully connected neural networks: the fist neural network approximates the low-fidelity data, while the second and third NNs are for constructing the correlations between the low- and high-fidelity data, and the last NN  encodes the PDEs that describe the corresponding physical problems. The two sub-networks included in the high-fidelity network are employed to approximate the linear and nonlinear parts of the correlations, respectively. Training the relaxation parameter employed to combine the two sub-networks enables the MPINNs to learn the correlation based on the training data without any prior assumption on the relation between the low- and high-fidelity data. 

MPINNs have the following attractive features: (1) Owing to the expressible capability of function approximation of the NNs, multi-fidelity NNs are able to approximate both continuous and discontinuous functions in high dimensions; (2) Due to the fact that NNs can handle almost any kind of nonlinearities, MPINNs are effective for identification of unknown parameters or functions in inverse problems described by nonlinear PDEs.

We first tested the new multi-fidelity DNN in approximating continuous and discontinuous functions with linear and nonlinear correlations.  Our results demonstrated that the present model can adaptively learn the correlations between the low- and high-fidelity data based on the training data
of variable fidelity. In addition, this model can easily be extended based on the embedding theory to learn more complicated nonlinear and non-functional correlations. We then tested MPINNs on inverse PDE problems, namely,  in estimating the hydraulic conductivity for unsaturated flows as well as the reaction models in reactive transport in porous media. We found that the proposed new MPINN can identify the unknown parameters or even functions with high accuracy  using very few high-fidelity data, which is  promising in reducing the high experimental cost for collecting high-fidelity data. Finally, we point out that MPINNs can also be employed for high-dimensional problems as well as problems with multiple fidelities, i.e. more than two fidelities.

\section*{Acknowledgement}
This work was supported by the PHILMS grant DE-SC0019453, the DOE-BER grant DE-SC0019434, the AFOSR grant FA9550-17-1-0013, and the DARPA-AIRA grant HR00111990025. In addition, we would like to thank Dr. Guofei Pang, Dr. Zhen Li, Dr. Zhiping Mao, and Ms Xiaoli Chen for their helpful discussions.

\begin{appendix}
\section{Data-driven manifold embeddings}
\label{embedding}
To learn more complicated non-linear correlation between the low- and high-fidelity data, we can further link the multi-fidelity DNNs with the embedding theory \cite{lee2018linking}. According to the weak Whitney embedding theorem \cite{whitney1936differentiable}, any continuous function from an $n$-dimensional manifold to an $m$-dimensional manifold may be approximated by a smooth embedding with $m > 2n$.  Using this theorem, Taken's theorem \cite{takens1981detecting} further points out that the $m$ embedding dimensions can be composed of $m$ different observations of the system state variables or $m$ time delays for a single scalar observable.

Now we will introduce the applications of the two theorems in multi-fidelity modelings. We assume that both $y_L$ and $y_H$ are smooth functions. Suppose that $y_L, y_L(x - \tau), ..., y_L(x - (m-1)\tau)$ ($\tau$ is the time delay)  and a small number of $(x, y_H)$ are available, we can then express $y_H$ in the following form 
\begin{align}
y_H(x) = \mathcal{F}(x, y_L(x), y_L(x - i \tau)), i = 1,...,m-1. 
\end{align}
By using this formulation, we can construct more complicate correlation than Eq. \eqref{nlinear}. To link the multi-fidelity DNN with the embedding theory, we can extend the inputs for $\mathcal{NN}_{H,i}$ to higher dimensions, i.e., $\left[ x, y_{L}(x) \right] \rightarrow [ x, y_{L}(x), y_L(x - \tau), y_L(x - 2\tau), ..., $ $ y_L(x - (m-1) \tau) ]$, which enables the multi-fidelity DNN to discover more complicated underlying correlations between the low- and high-fidelity data.

Note that the selection of optimal value for the time delay $\tau$ is important in embedding theory \cite{hegger1999practical,abarbanel1993analysis,dhir2017bayesian}, on which numerous studies have been carried out \cite{hegger1999practical}. However, most of the existing methods for determining the optimal value of $\tau$ appear to be problem-dependent \cite{hegger1999practical}. Recently, Dhir {\sl et al.} proposed a Bayesian delay embedding method, where $\tau$ is robustly learned from the training data by employing the variational autoencoder \cite{dhir2017bayesian}.  In the present study, the value of $\tau$ is also learned by optimizing the NNs rather than setting it as constant as in the original work presented in Ref. \cite{lee2018linking}.

\end{appendix}





\bibliographystyle{elsarticle-num}
\biboptions{sort&compress}
\bibliography{mpinns}







\end{document}